\documentclass[referee]{aa}
\usepackage{longtable}
\usepackage{supertabular}
\usepackage{graphics}
\usepackage{natbib}
\begin{document}
\thesaurus{0.8(0.9.10.1; 0.9.13.2; 0.8.16.5)}
\title{ROSAT-SDSS Galaxy Clusters Survey.} 
\subtitle{I. The Catalog and the correlation of X-ray and optical properties.}
\author{P. Popesso\inst{1}, H. B\"ohringer\inst{1}, J. Brinkmann\inst{2}, W. Voges\inst{1}, D. G. York\inst{3}}
\institute{ Max-Planck-Institut fur extraterrestrische Physik, 85748 Garching, Germany
\and Apache Point Observatory, P. O. Box 59, Sunspot, NM 88349
\and Department of Astronomy and Astrophysics, University of Chicago,5640 South Ellis Avenue, Chicago, IL 60637}
\authorrunning{P. Popesso et al.}
\maketitle

\begin{abstract}
For a detailed comparison of the appearance of  cluster of galaxies in
X-rays and  in the optical,  we have compiled a comprehensive database
of X-ray  and optical properties of a  sample of clusters based on the
largest available X-ray and optical  surveys: the ROSAT All Sky Survey
(RASS) and  the Sloan Digital  Sky Survey  (SDSS).  The  X-ray  galaxy
clusters of this RASS-SDSS catalog cover a  wide range of masses, from
groups of
 $10^{12.5}$ $M_{\odot}$  to  massive clusters of  $10^{15}$
$M_{\odot}$
 in the redshift range  from 0.002 to 0.45.  The RASS-SDSS
sample comprises  all the X-ray selected   objects already observed by
the Sloan Digital Sky
 Survey (114 clusters).  For each system we have
uniformly determined the X-ray    (luminosity  in the  ROSAT    band,
bolometric luminosity,   center  coordinates)  and  optical properties
(Schechter luminosity function
 parameters, luminosity, central galaxy
density, core, total  and
 half-light radii).  For  a  subsample of 53
clusters we have also compiled the  temperature and the iron abundance
from the literature.  The  total optical luminosity can be  determined
with a  typical uncertainty of 20\%  with a  result independent of the
choice  of local  or  global background subtraction.   We searched for
parameters which  provide   the  best correlation between   the  X-ray
luminosity and  the optical properties   and found  that  the z   band
luminosity  determined within  a   cluster aperture  of    0.5 Mpc
$\rm{h}_{70}^{-1}$   provides the best  correlation with a scatter of
about 60-70\%.  The scatter  decreases to less  than 40\%  if the
correlation  is limited to the  bright X-ray clusters.  The resulting
correlation of $L_X$  and $L_{op}$   in  the z and   i  bands shows  a
logarithmic slope of 0.38, a value  not consistent with the assumption
of a constant $M/L$.  Consistency is found, however, for an increasing
M/L  with  luminosity as  suggested   by other observations.  We  also
investigated     the  correlation between  $L_{op}$     and the  X-ray
temperature obtaining the same result.
\end{abstract}

\section{Introduction}
Cluster  of galaxies are the largest  well  defined building blocks of
our
 Universe. They form via  gravitational collapse of cosmic
 matter
over a region of several megaparsecs. Cosmic baryons, which
 represent
approximately  10-15\%  of the  mass content of  the Universe,
 follow
dynamically the dominant dark matter during the collapse. As a
 result
of  adiabatic  compression and   of  shocks generated  by  supersonic
motions,  a    thin hot  gas   permeates   the  cluster gravitational
potential.  For a typical cluster  mass  of $10^{14}$ $M_{\odot}$ the
intracluster gas reaches a temperature of the order of $10^7$ keV and,
thus, radiates   optically  thin    thermal bremsstrahlung and    line
radiation in  the X-ray band.  In 1978, the launch  of the first X-ray
imaging telescope, the \emph{Einstein} observatory, began a new era of
cluster discovery, as clusters proved to be luminous ($\ge 10^{42-45}$
ergs $\rm{s}^{-1}$),  extended  ($\rm{r}\sim 1-5$ Mpc) X-ray  sources,
readily identified in the  X-ray sky. Therefore, X-ray observations of
galaxy clusters  provide an efficient  and physically motivated method
of identification  of  these structures.  The X-ray  selection is more
robust against contamination along the  line of sight than traditional
optical methods since  the richest  clusters  are relatively  rare and
since    X-ray  emission, which is    proportional  to the gas density
squared, is  far more sensitive to physical  overdensities than in the
projected  number   density of galaxies   in   the sky.  In  fact  the
existence  of diffuse,   very hot  X-ray  emitting  gas   implies  the
existence  of  a   massive confining   dark  matter  halo.  Moreover,
selection  according to X-ray luminosity is  also an efficient way to
find the highest mass  concentrations due to well
 defined correlation
between the X-ray luminosity and  the total cluster
 mass (Reiprich \&
Bh\"oringher.  2002). In addition  to allowing the
 identification  of
galaxy clusters, X-ray observations provide  a wealth
 of  information
on the intracluster medium  itself, e.g. its metal
 abundance,  radial
density    distribution and   temperature   profile.
   These   latter
quantities, in turn, can be used to reliably estimate 
 the
\emph{total mass} of the system.

In addition  to  the   hot,  diffuse  component,   baryons   are also
concentrated in the individual galaxies within the cluster. These are
best studied through photometric  and spectroscopic  optical surveys,
which provide  essential   information about luminosity,   morphology,
stellar
 population and  age. Solid observational evidences indicate a
strong
 interaction between  the two baryonic  components, as galaxies
pollute the
 intracluster medium  expelling metals via  galactic winds
producing the  observed  metal abundances  in clusters  (De Grandi  et
al. 2002, Finoguenov et al. 2001). On the other hand, the evolution of
galaxies in clusters is  influenced by processes  due  to the hot  gas
(e.g. gas stripping  by  ram pressure, etc.)   as  it is by  internal
processes like star formation,  galactic winds, supernovae  explosions
etc., operating  inside galaxies themselves   (Dressler et al.   1997,
Fasano et al. 2000, van Dokkum et al.  2000, Lubin et al. 2002, Kelson
et  al.  1997,2000, Ziegler\&Bender   1997,  Gomez et  al.  2003).  In
conclusion, understanding the complex  physics at play
 in clusters of
galaxies   requires combined   X-ray  and  optical   observation  of a
statistically significant sample of these objects.

On  the    basis of  these considerations,  we    have created a large
database
 of clusters of galaxies based on the largest available X-ray
and
 optical surveys: the ROSAT All Sky Survey  (RASS), the only X-ray
all
 sky survey ever  realized using an  imaging X-ray  telescope, and
the Sloan  Digital Sky Survey  (SDSS),
 which  is observing the  whole
Northern Galactic  Cap and part of  the
 Southern Galactic Cap in five
wide  optical bands covering the entire
  optical range.  By carefully
combining  the data of the two  surveys we
 have created the RASS-SDSS
galaxy cluster catalog.  Although two galaxy cluster catalogs from the
SDSS already exist, the Cut and Enhance Galaxy Cluster Catalog of Goto
et al.  (2002)  and  the  Merged Cluster   Catalog of  Bahcall et  al.
(2003, see also Kim et al. 2002), we prefered to compile a new cluster
catalog by selecting the systems in the  X-ray band, for which we have
reliable X-ray  characteristics and  for the reasons  explained above.
The X-ray-selected
 galaxy clusters cover a wide range of masses, from
groups of
 $10^{12.5}$ $M_{\odot}$  to  massive clusters of  $10^{15}$
$M_{\odot}$
  in a redshift range  from 0.002  to 0.45.  The RASS-SDSS
sample comprises  all the X-ray  detected objects already  observed in
the sky region covered by the Sloan Digital Sky
 Survey.

One of the first goals is the comparison  of the X-ray and the optical
appearance  of   the clusters.   We want   in  particular find optical
parameters that  provide the    closest   correlation to the     X-ray
parameters, such that we can predict  within narrow uncertainty limits
the X-ray luminosity from these  optical parameters and vice versa.  So
far  optical     and   X-ray cluster surveys    have    been conducted
independently without much  intercomparison.  Therefore, the empirical
relationship between the   X-ray luminosity and optical  luminosity of
clusters is   not so  well defined,  in   large part  because  of  the
difficulties inherent in  measuring the cluster optical luminosity and
in getting a homogeneous set of total optical luminosities for a large
number of clusters. Edge and Stewart  (1991) found that the bolometric
X-ray  luminosity  of  a    local sample of    X-ray-selected clusters
correlated very roughly with Abell number and somewhat better with the
Bahcall galaxy   density (number    of  bright galaxies    within  0.5
$\rm{h}^{-1}$ Mpc; Bahcall 1977,1981) for the small subsample that had
Bahcall galaxy densities.  Arnaud et al.  (1992) made an heroic effort
in computing   cluster  optical luminosities at   low redshift  from a
heterogenous literature.  The    first joint X-ray/optical survey   of
galaxy clusters was  the ROSAT Optical X-ray  Survey (ROXS, Donahue et
al.  2002).  They observed 23 ROSAT pointings for a  total of 5 square
degrees in the I band and partially in the V band. The X-ray selection
and    the optical   selection    of  cluster   candidates were   done
independently, with the wavelets algorithm in the former case and with
a matched  filter algorithm in the  latter one.   They found X-ray and
optical coincident detections for 26  galaxy clusters.  Donahue et al.
(2001)  studied  the relation  between  the  X-ray luminosity  and the
matched  filter  parameter  $\Lambda   _{cl}$, which  is approximately
equivalent to the  number of $L^*$ galaxies  in the system (Postman et
al.  1996).  They  found a marginally significant correlation  between
the  two quantities with  a  significant scatter.   Yee and  Ellingson
(2003) defined  a  new  richness parameter  as  the  number of cluster
galaxies  within some fixed  aperture, scaled by a luminosity function
and  a  spatial distribution function.  They   analysed a sample of 15
clusters from  CNOC1  Cluster Redshift  Survey, and  found a very poor
correlation between this  parameter and other cluster  properties such
as the X-ray luminosity, temperature and the velocity dispersion.

In the present  paper we describe  the properties  and the information
contained in the RASS-SDSS   catalog  and the  resulting  correlations
between the X-ray and  optical properties in  the sample. In section 2
we explain how the cluster sample  has been created by X-ray selecting
the systems from  the available X-ray  cluster and group  catalogs. In
section  3 we describe the   method for calculating  the X-ray cluster
properties. We describe in section 4 the optical data and in section 5
the  data  reduction method. We  analyse and  discuss the correlations
between the optical luminosity and the X-ray  properties in section 6.
We summarize and   discuss the catalog properties  and  the results in
section 7.

For    all     derived   quantities,   we   have    used $\rm{H}_0=70$
$\rm{km}$$\rm{s}^{-1}$$\rm{Mpc}^{-1}$, $\Omega_ {m}=0.3$  and $\Omega_
{\lambda}=0.7$.

\section{The construction of the sample.}
In order  to correlate optical and X-ray  galaxy cluster properties it
is necessary to have a large statistical sample and to cover the whole
mass range of  the systems considered.  Since  the X-ray observations
provide a robust  method of identification of  galaxy clusters and the
X-ray  luminosity is  a  good estimator of  the  system total mass, in
principle the best approach should  be constructing a cluster  catalog
of X-ray selected objects in a wide range  of X-ray luminosity.  While
for intermediate and high X-ray   luminosity (mass), several  complete
catalogs of X-ray selected clusters  already exist (NORAS and REFLEX),
in the low mass range a systematic X-ray survey of groups has still to
be  carried out.  As  a consequence,  it is  impossible  at the moment
constructing a strictly X-ray selected cluster  sample, which covers a
wide range of   masses from very poor groups    to rich clusters.    A
reasonable compromise  in order  to fill the   low  mass range of  the
spectrum  is to select  all the low  mass clusters and  groups to date
X-ray detected,  even if  they  are selected in  other wavebands. This
compromise is  acceptable for our  purposes, since  we  do not want to
carry out an X-ray survey of galaxy clusters, but a systematic analysis
of  the correlation  between  X-ray  and  optical  properties of those
systems.

By following  these criteria, the  intermediate mass  and high
mass  clusters  have been  selected   from three  ROSAT based  cluster
samples: the ROSAT-ESO  flux   limited X-ray cluster sample   (REFLEX,
B\"ohringer et  al.  2003), the  Northern ROSAT All-sky cluster sample
(NORAS,  B\"ohringer et  al. 2000)  and  the  NORAS 2  cluster  sample
(Retzlaff 2001) .  REFLEX is a complete sample of clusters detected in
13924 $\rm{deg}^2$ in the southern hemisphere down  to a flux limit of
3 $10^{-12}$ $\rm{erg} \rm{s}^{-1}
\rm{cm}^{-2}$ in the $0.1-2.4$ $\rm{keV}$ comprising 448 clusters. The
NORAS  galaxy cluster survey   contains 495 clusters  showing extended
emission   in the RASS  in the  northern   hemisphere with count rates
$C_X\ge 0.06$ $\rm{counts}$ $\rm{s}^{-1}$  in the $0.1-2.4$ keV. NORAS
2 is  the continuation of   the NORAS project and  aims  at a complete
survey of X-ray galaxy clusters, in 13598 $\rm{deg}^2$ of the northern
hemisphere,  down  to  a   flux limit  of   $2$  $10^{-12}$  $\rm{erg}$
$\rm{s}^{-1} \rm{cm}^{-2}$ in the same X-ray band, which gives rise to
an expected total of  about 800 clusters. The  samples are based on an
MPE    internal X-ray source    catalog  extracted  with a   detection
likelihood $\ge 7$.

The low  mass  clusters and the   groups have been selected   from two
catalogs of X-ray detected   objects: the ASCA Cluster Catalog   (ACC)
from Horner et al. (2001) and the Group Sample (GS) of Mulchaey et al.
2003. The ACC  is a  collection of  all the clusters  retrieved in the
ASCA archive and  discovered with different  selection strategies.  It
contains measured   luminosities,   average temperatures,   and  metal
abundances  for 273  clusters and groups.   The GS  is a heterogeneous
collection  of 66 ROSAT   systems, especially optically selected, with
velocity dispersions   less than  600   $\rm{km}$ $\rm{s}^{-1}$  or an
intragroup medium temperature less than 2 $keV$.

The RASS-SDSS  galaxy cluster sample  comprises all the X-ray clusters
of the  cited catalogs in the  area covered by  the  Sloan Digital Sky
Survey up to  February 2003.  For each X-ray    system in the   common
RASS-SDSS area  we found an optical  counterpart.  The sample includes
114 galaxy clusters, 14 of  which come from  REFLEX, 72 from NORAS and
NORAS 2, 8  from  Mulchaey's  groups sample   and 20 from   ACC.   The
RASS-SDSS  galaxy cluster   sample, therefore, can  not be  considered
strictly an X-ray 'selected'  cluster  sample, but should  be  defined
more   precisely  an    X-ray  'detected'  cluster  sample  for    the
heterogeneous selection of the low mass range systems.

\section{The X-ray data.}
In order to create a homogeneous catalog  of X-ray cluster properties,
we have calculated all X-ray  parameters using only  RASS data for all
clusters in the sample.  The X-ray luminosity has been calculated with
the growth curve  analysis (GCA) method used  for REFLEX and  NORAS 2,
based on the RASS3  database (Voges et  al. 1999).  The GCA method was
applied to RASS3 at  the position of  the clusters. The method allowed
for a  small adjustment of   the  position to   center  on the   X-ray
maximum.    Since  within  the GCA   aperture    the measured flux  is
underestimated typically by  an amount   of $7-10\%$ (B\"ohringer   et
al. 2000)   the  missing flux    is estimated  by  assuming  an  X-ray
luminosity   scaled cluster  model   (B\"ohringer   et al. 2001)   and
corrected  for. A first approximate unabsorbed  flux is calculated for
each X-ray source from the observed  count rate, by assuming a thermal
spectrum with  a temperature of 5 keV  and a metallicity  of 0.3 solar
and without  a K-correction.  Then  an iterative computation  uses the
redshift and the unabsorbed X-ray flux to give a first estimate of the
luminosity. With   the luminosity-temperature  relation  of Markevitch
(1998,  without  correction for cooling  flows)   a better temperature
estimate  is obtained,  and  the count  rate-flux conversion factor is
recomputed including  the appropriate  K-correction for the  redshift,
resulting in a  corrected flux and a new  X-ray  luminosity. The X-ray
luminosities  as  used  in this  paper  are   calculated  in the ROSAT
$(0.1-2.4)$ keV  energy band in the  cluster rest-frame for a $\Lambda$
cosmology with the  parameter given  above.  The GCA also  returns for
each source many  physical  parameters like improved  source position,
background brightness, spectral hardness ratio, and KS probability for
source extent.

The X-ray bolometric  luminosity   has been  derived from  the   X-ray
luminosity  in  the  ROSAT  $(0.1-2.4)$ keV   energy   band. A first
estimation  of  the cluster   temperature  is calculated  by using the
$L_X-T_X$  relation of   Xue \&  Wu  2000  in order   to  estimate the
appropriate bolometric correction.

For a subsample of 53 galaxy  clusters we have also compiled from
the literature the ASCA and XMM temperature  and iron abundance of the
intracluster medium in the system.

\section{Optical data}

\begin{figure}
\begin{center}
\begin{minipage}{1.0\textwidth}
\resizebox{\hsize}{!}{\includegraphics{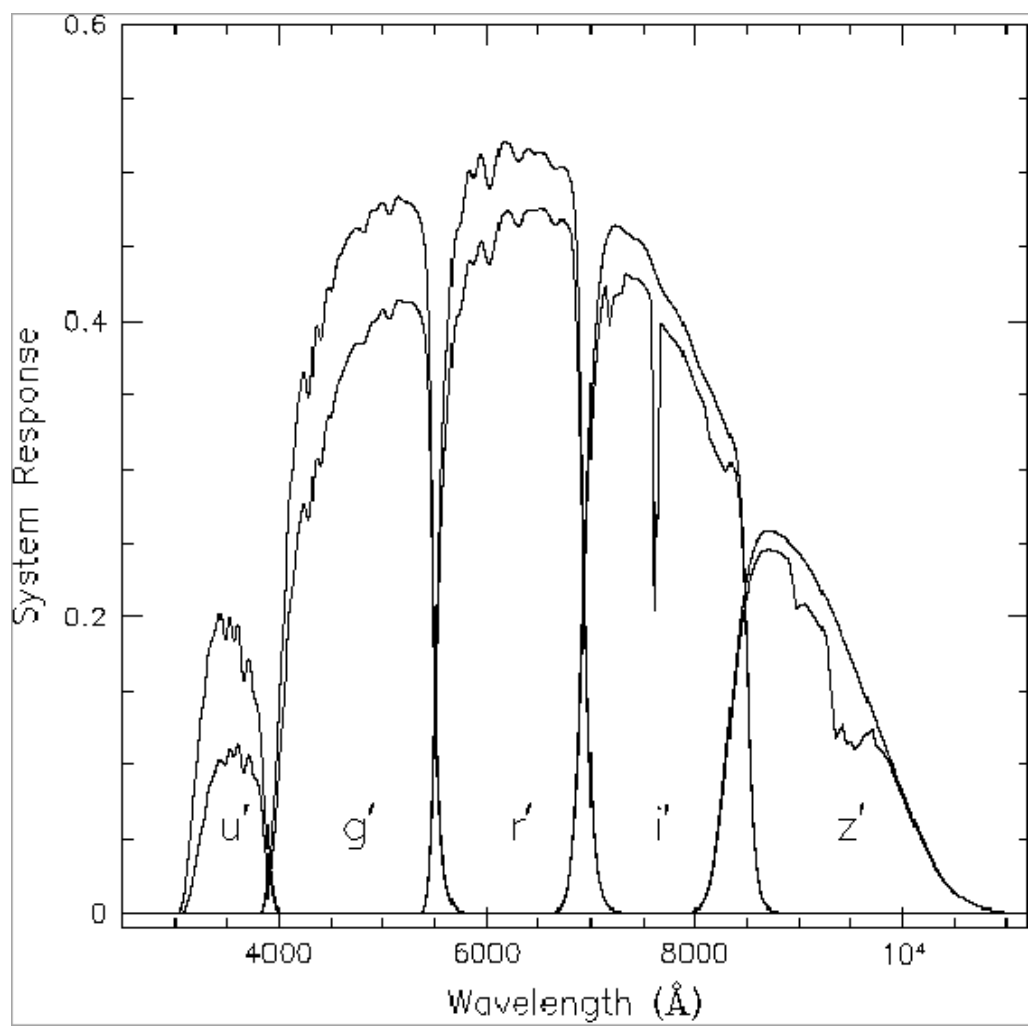}}
\end{minipage}
\end{center}
\caption{
Response  function  of  the  SDSS photometric  system.   Dashed curves
indicate  the  response  function including atmospheric transmission at
1.2 airmass at the altitude of Apache Point Observatory.}
\label{filter}
\end{figure}

The    optical  photometric data were taken     from the SDSS (York et
al. 2000 and Stoughton et al.  2002). The SDSS  consists of an imaging
survey of  $\pi$ steradians of the northern  sky in the five passbands
u,  g, r ,i,  z, in   the  entire optical range  from  the atmospheric
ultraviolet cutoff in the blue to the  sensitivity limit of silicon in
the red (Fig. \ref{filter}).  The survey is carried  out using a 2.5 m
telescope, an imaging  mosaic   camera with  30 CCDs,  two   fiber-fed
spectrographs and a 0.5 m  telescope for the photometric  calibration.
The imaging survey is taken in drift-scan mode.   The imaging data are
processed  with a photometric   pipeline (PHOTO) specially written for
the SDSS data.

\subsection{The galaxy sample}
In order  to study the  optical cluster properties,  we have created a
complete galaxy sample for each cluster,  by selecting the galaxies in
an area centered on the  X-ray source position  and with radius of 1.5
deg.  We used the selection criteria of Yasuda et al. (2001) to define
our  galaxy sample from the  photometric catalog produced by PHOTO. We
have  selected only objects  flagged  with PRIMARY,  in order to avoid
multiple detections in the overlap between  adjacent scan lines in two
strips of a stripe and  between adjacent frames. Objects with multiple
peaks (parent) are  divided by the  deblender in different  components
(children); if the objects can not be deblended, they are additionally
flagged   as  NODEBLEND.   Only isolated   objects,  child objects and
NODEBLEND flagged objects are  used in constructing our galaxy sample.
The star-galaxy  separation    is performed in   each   band using  an
empirical technique, based on the difference between the model and the
PSF  magnitude.  An   object is classified  as  galaxy  if the   model
magnitude and the Point Spread Function (PSF) magnitude differ by more
than 0.145. This method  seems to be robust  for $ r\le 21$ mag, which
is also the completeness limit of the survey  in the Northern galactic
cap.  Since saturated pixels and diffraction spikes can compromise the
model-fitting algorithm, some stars  can be misclassified as galaxies.
Therefore we have rejected all  object with saturated pixels and which
are flagged as BRIGHT.  Furthermore, we have classified an object as a
galaxy only if PHOTO classified it as a galaxy  in at least two of the
three  photometric  bands g, r, i.    After a  visual  inspection of a
sample of galaxies with $r\le 16$ mag, Yasuda et al.  (2001) concluded
that  the described selection criteria  give a  sample completeness of
97\%, and the same completeness is found for the sample at $16 \le r
\le 21$ mag after comparison with the Medium Deep Survey catalog (MDS)
constructed using WFPC2 parallel images from  HST.  The major
reason of missing real galaxies is the rejection of galaxies blended
with saturated stars, while spurious galaxy detections are double
stars or shredded galaxies with substructures.

\subsection{SDSS Galaxy photometry}
Since   the galaxies  do  not have  sharp   edges or a  unique surface
brightness profile, it is nontrivial to define a flux for each object.
PHOTO  calculates a number  of different  magnitudes  for each object:
model magnitude, Petrosian magnitudes  and  PSF magnitudes. The  model
magnitudes  are calculated  by fitting de  Vaucouleurs and exponential
model, convolved with the local PSF, to the  two dimensional images of
the  galaxies in the r band.  The  total magnitude are determined from
the better  fitting of   the  two shape  function. Galaxy   colors are
measured by applying the best fit model of an object in  the r band to
the  others bands, measuring the flux  in the same effective aperture.
However due to a bug of PHOTO, found during the completion of DR1, the
model magnitudes  are systematically    under-estimated by about   0.2
magnitudes for galaxies brighter then  20th magnitude, and accordingly
the        measured      radii  are     systematically     too   large
(http://www.sdss.org/DR1/products/catalogs/index.html).   This   error
does  not affect  the   galaxy colors  but  makes the  model magnitude
useless for the determination of the galaxy total luminosities.

The Petrosian flux is defined by
\begin{equation}
F_p=2 \pi \int^{f_2r_P}_{0}{I(r)dr} ,
\end{equation}
where $I(r)$ is the surface brightness profile of the galaxy, and
$r_P$ is the Petrosian radius satisfying the equation:
\begin{equation}
f_1=\frac{2 \pi \int^{1.25r_P}_{0.8r_P}{\frac{I(r)rdr}{{\pi
[(1.25r_P)^2-(0.8r_P)^2]}}}}{2 \pi \int^{r_P}_0{\frac{I(r)rdr}{(\pi
r_p^2)}}}.
\end{equation}
The Petrosian aperture is set to $2r_P$, and  it encompasses 99$\%$ of
the galaxy total light in case of an exponential profile and 82$\%$ in
case of a de Vaucouleurs profile (Blanton  et al 2001).  The Petrosian
ratio  $f_1$ is set to 0.2;  at smaller values  PHOTO fails to measure
the Petrosian  ratio, since the $S/N$  is too low.   For faint objects
the effect of the seeing on Petrosian magnitude is not negligible.  As
the galaxy size becomes comparable to the seeing disk, the fraction of
light measured by the Petrosian quantities approaches the fraction for
a PSF, about  $95\%$, in which case the  flux is reduced  for a galaxy
with  an exponential  profile  and increased for   a galaxy  with a de
Vaucouleurs   profile (Strauss  et    al. 2002).  Thus  the  Petrosian
magnitudes  are the  best   measure  of the   total light  for  bright
galaxies, but fail to  be a  good measure for  faint objects.   In the
data analysis of   this paper  we  used the   Petrosian magnitudes for
galaxies brighter then   20  mag and  the psf  magnitudes  for objects
fainter than 20 mag.

\section{Optical Luminosity from SDSS data}

\subsection{Background subtraction}
The total optical luminosity  of a cluster has  to be calculated after
the  subtraction    of   the    foreground   and   background   galaxy
contamination. Since we have used only photometric  data from the SDSS
galaxy  catalog, we have  no direct information  on the cluster galaxy
memberships.   There  are two different  approaches  to  overcome this
problem. Since galaxy clusters show a very well define red sequence in
the color magnitude diagram, a galaxy color cut could be use to define
the cluster membership (Gladders et  al. 2000). On  the other hand the
background  subtraction  can  be based  on  the  number counts of  the
projected field galaxies  outside  the cluster.  We  chose  the latter
approach since the former method  may introduce  a bias against  bluer
cluster  members  observed to  have    varying  fractions due  to  the
Butcher-Oemler effect.

\begin{figure}
\begin{center}
\begin{minipage}{1.0\textwidth}
\resizebox{\hsize}{!}{\includegraphics{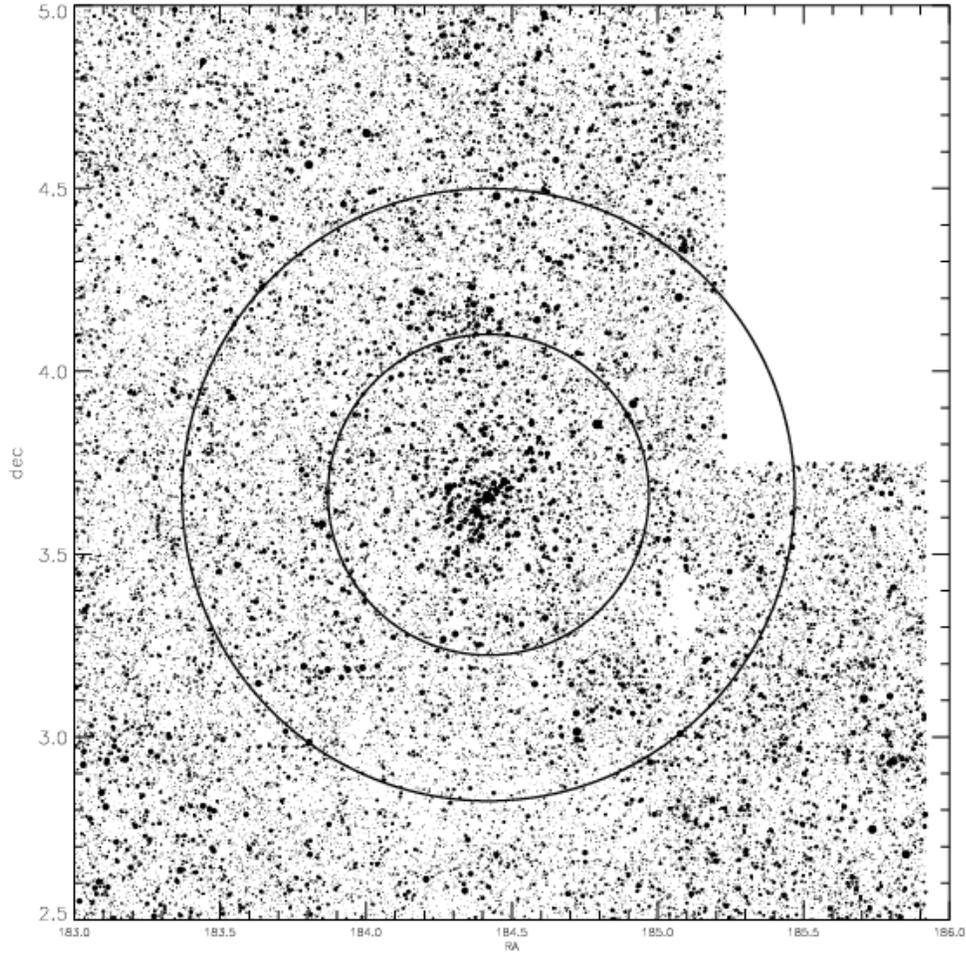}}
\end{minipage}
\end{center}
\caption{
The plot shows  a cluster region  with the local background.  The dots
represent the galaxies in the sample.  The  biggest dots correspond to
the brightest galaxies in  apparent r magnitude. The  local background
number  counts have  been  calculated inside  the   annulus with inner
radius equal  to $r_{200}+0.2$ deg and  a  width of $0.5$  degree. The
regions with voids due  to lack of data or  close clusters have to  be
discarded in the background estimation.}
\label{local}
\end{figure}

We have considered   two   different approaches to  the    statistical
subtraction  of the  galaxy  background.  First  we have  calculated a
local background. The $M_{200}-L_X$ relation of Reiprich  et al.  2002
was used  to  compute the   $r_{200}$ radius  (where the cluster  mass
density is 200 times the critical cosmic mass density), as a pragmatic
approximation   of the virial radius.    Then,  we defined an  annulus
centered on  the cluster X-ray center, with  an inner  radius equal to
$r_{200}+0.2$ deg  and a width of  $0.5$ degree (Fig. \ref{local}). In
this way  the galaxy  background  has been estimated well  outside the
cluster but still  locally.  The annulus  has been then  divided in 20
sectors (  analogous to the approach  in B\"ohringer  et al. 2001) and
those  featuring a  larger than  $3\sigma$ deviation  from  the median
galaxy density  are discarded from  the further  calculation.  In this
way other clusters close to  the target or voids  are not included  in
the background correction.    We  have computed  the   galaxies number
counts $N_{bg}^l(m)dm$ per bin of  magnitude (with a bin width of 0.5
mag) and per squared degree in the remaining area of the annulus.  The
statistical  source  of  error  in this   approach   is the Poissonian
uncertainty of the counts, given by $\sqrt{(N_{bg}^l(m))}$.

\begin{figure}
\begin{center}
\begin{minipage}{1.0\textwidth}
\resizebox{\hsize}{!}{\includegraphics{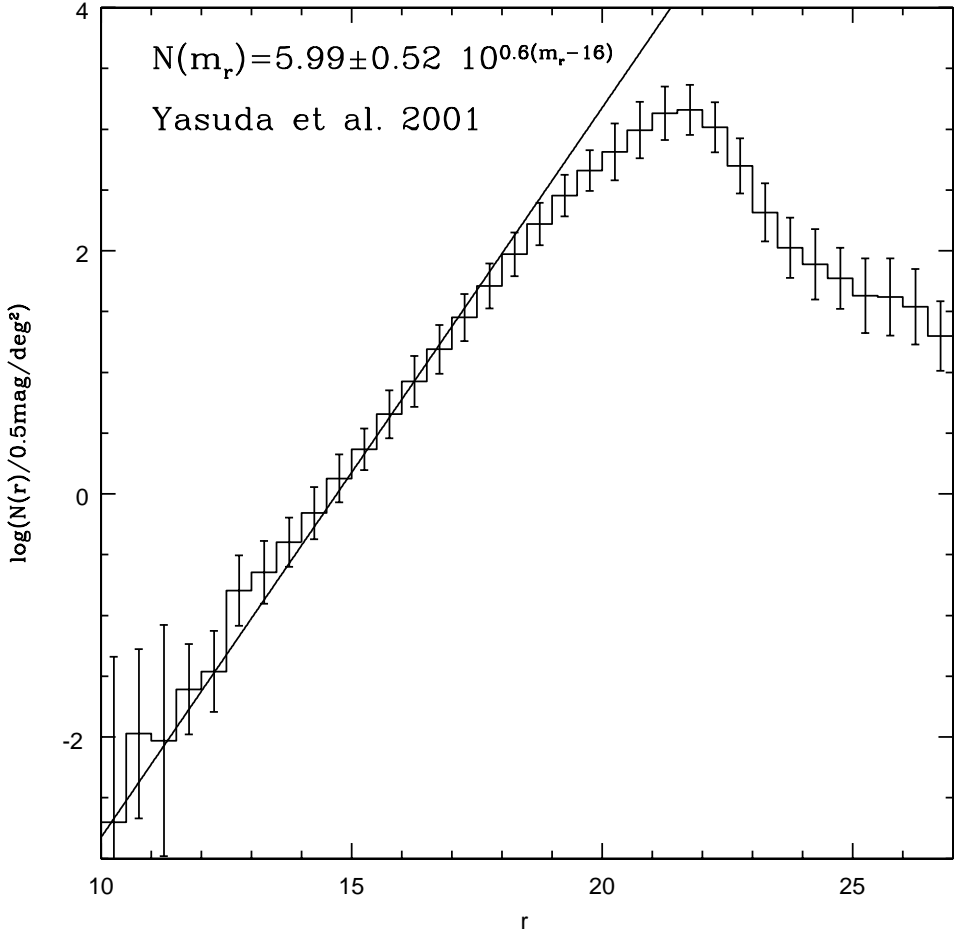}}
\end{minipage}
\end{center}
\caption{
Global background number counts as a function  of magnitude in the $r$
band. The  error  bars  include   the  contribution of   large   scale
structure. The line shows the counts-magnitude relation expected for a
homogeneous galaxy distribution in a universe with Euclidean geometry:
$N(r)=A_r10^{0.6(r-16)}$.           The          value              of
$A_r=5.99\pm0.52(0.5mag)^{-1}deg^{-2}$  is the  results  of the fit in
Yasuda et al. 2001 for $12\le r \le 17$.}
\label{global}
\end{figure}

As a second method we have derived a global background correction. The
galaxy number counts $N_{bg}^g(m)dm$ was  derived from the mean of the
magnitude number counts determined in five different SDSS sky regions,
each  with an area of  30 $\rm{deg^2}$ (Fig.\ref{global}).  The source
of uncertainty  in this second case  is  systematic and originates the
presence of large-scale clustering within the galaxy sample, while the
Poissonian error  of the galaxy counts is  small due to the large area
involved. We have  estimated this error as  the standard deviation  of
the mean global number  counts, $\sigma_{bg}^g(m)$, in  the comparison
of  the  five areas.  In  order to  take into account  this systematic
source of error also for the the local background  , we have estimated
the           background      number      counts        error       as
$\sigma_{bg}(m)=max(\sqrt{(N_{bg}^l(m))},\sigma_{bg}^g(m))$   (Lumdsen
et al. 1997) for all the derived quantities.

After the background subtraction we found that the  signal to noise in
the u band  was too low to be  useful, and performed our  analysis on
the 4 remaining Sloan photometric bands g,r,i,z.

\subsection{Luminosity Function}
For each cluster and in all photometric bands we  have assumed that the
distribution function of galaxies in magnitude can be described by the
Schechter Luminosity function (LF):
\begin{equation}
\phi(m)dm=0.4~ln(10)~N_{clu}~\phi^*~10^{-0.4~(m-m^*)~(\alpha-1)}~exp(-10^{-0.4~(m-m^*)})~dm.
\end{equation}
In the  equation $N_{clu}$ is  the number of  cluster galaxies and was
computed as the difference between the total number of galaxies in the
cluster region and the expected number  of interlopers, estimated from
the  local  (global)  background  galaxy   density.  $\phi^*$ is   the
normalization of the  Schechter   Luminosity function, given   by  the
inverse  of   the integral of  the  LF  over the  considered magnitude
range.  To determine  the remaining  parameters  $M^*$ and $\alpha$ we
fitted the Schechter  LF to the data with  a Maximum Likelihood Method
(MLM, Sarazin  1980). Since we  have no information  about the cluster
membership,  we  have   considered   the  observed   galaxy  magnitude
distribution   in  the cluster  region as   the   sum of the Schechter
Luminosity  function  ($\phi(m)$)    plus a  background   contribution
($b(m)$):
\begin{equation}
\Phi(m)=\frac{\phi(m) + b(m)}{N_{tot}}.
\end{equation}
$\Phi(m)$ is normalized to unity when integrated over the
considered range of magnitudes.  

In order to perform a ML analysis, the  background contribution has to
be specified at any magnitude.  To estimate the $b(m)$, one can try to
fit the background number counts by a model. While the behavior of the
$N(m)-m$   relation   is  well  know    at the  bright   end   (in the
$logN(m)-log m$ is a line with slope 0.6, Fig.\ref{global}) , it is not
well understood at  the faint  end  (Yasuda et  al. 2001, Lumdsen   et
al. 1997).  Therefore instead of using a  specific functional form, we
simply have used a spline  to interpolate the background galaxy number
counts and estimated $b(m)$ at any magnitude.

\begin{figure}
\begin{center}
\begin{minipage}{1.0\textwidth}
\resizebox{\hsize}{!}{\includegraphics{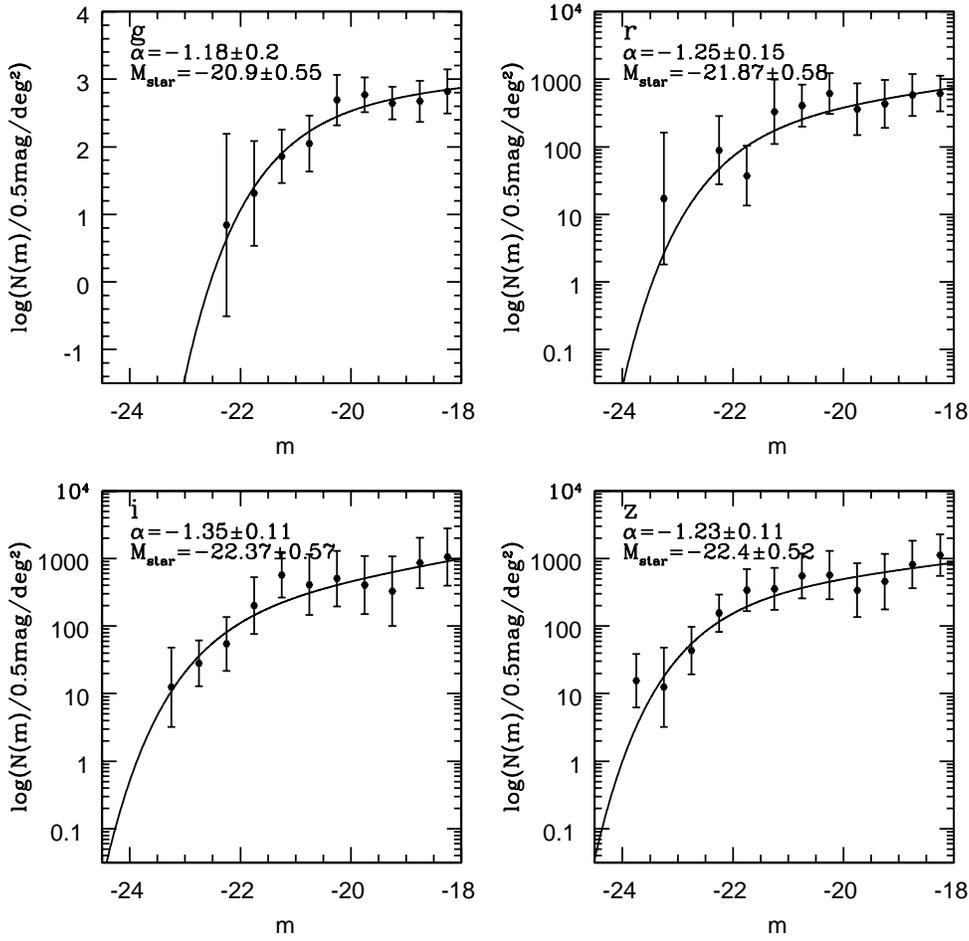}}
\end{minipage}
\end{center}
\caption{
Example of the individual fitted luminosity function of a cluster in 4
Sloan  photometric  bands. The   value  of the  fitted  parameters are
indicated in each panel.}
\label{lf}
\end{figure}

\begin{figure}
\begin{center}
\begin{minipage}{1.0\textwidth}
\resizebox{\hsize}{!}{\includegraphics{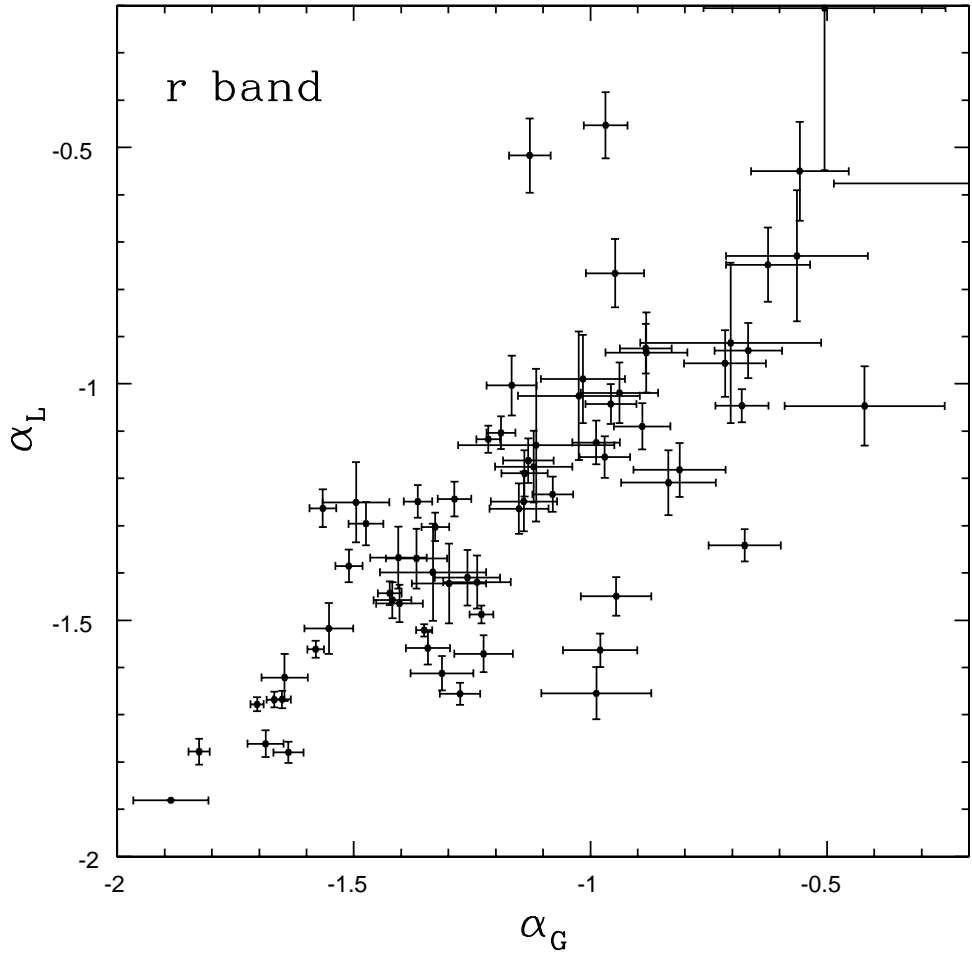}}
\end{minipage}
\end{center}
\caption{
Comparison of   the $\alpha$  parameter   (slope) of   the  individual
Schechter luminosity  functions. $\alpha _{L}$  is calculated by using
the local background correction, while  $\alpha _{G}$ is the result of
the global background correction. The different background subtractions
give consistent  results.  The  error bars  in  the  plot are  at 68\%
confidence level.}
\label{alpha}
\end{figure}

\begin{figure}
\begin{center}
\begin{minipage}{1.0\textwidth}
\resizebox{\hsize}{!}{\includegraphics{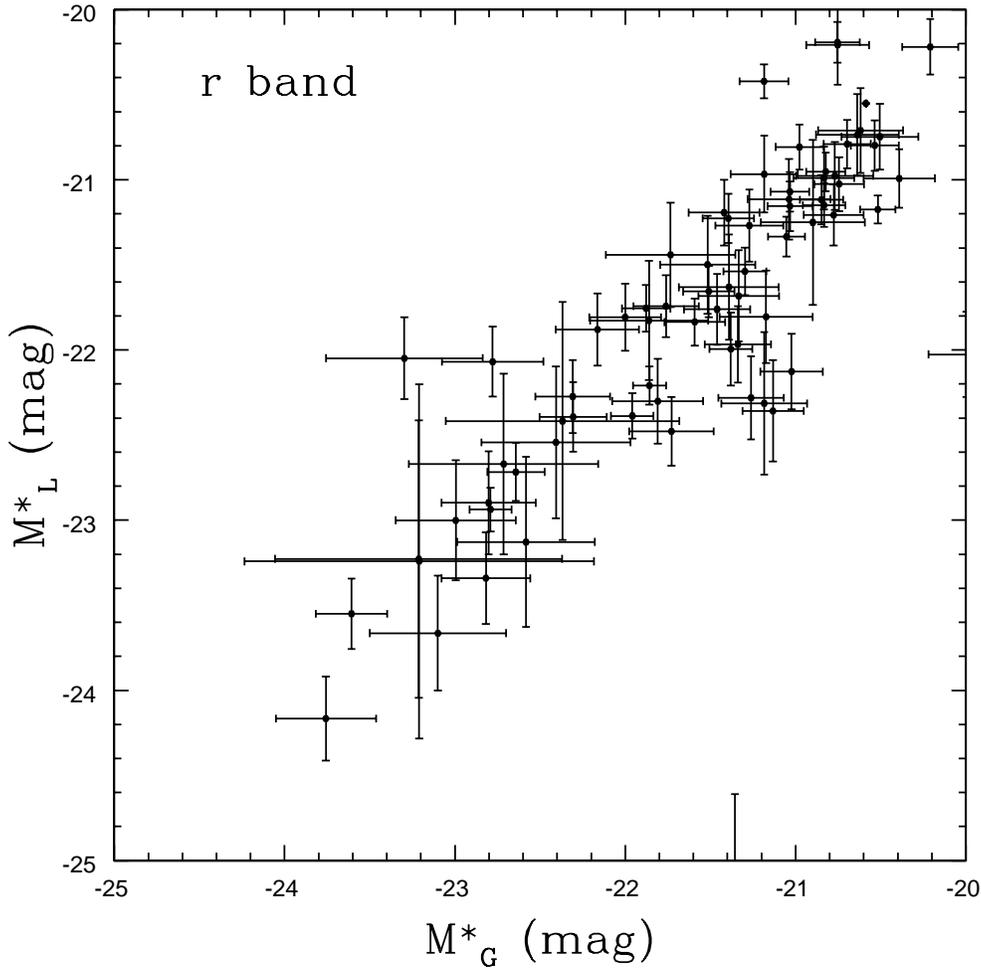}}
\end{minipage}
\end{center}
\caption{
Comparison  of the $M*$ parameter (knee)   of the individual Schechter
luminosity  functions.  $M*_L$   is  calculated  by using    the local
background  correction,  while  $M*_G$  is  the  result of  the global
background correction.   As in the   case of the  slope, the different
background subtractions give consistent results.  The error bars in the
plot are at 68\% confidence level.}
\label{mc}
\end{figure}

\begin{figure}
\begin{center}
\begin{minipage}{1.0\textwidth}
\resizebox{\hsize}{!}{\includegraphics{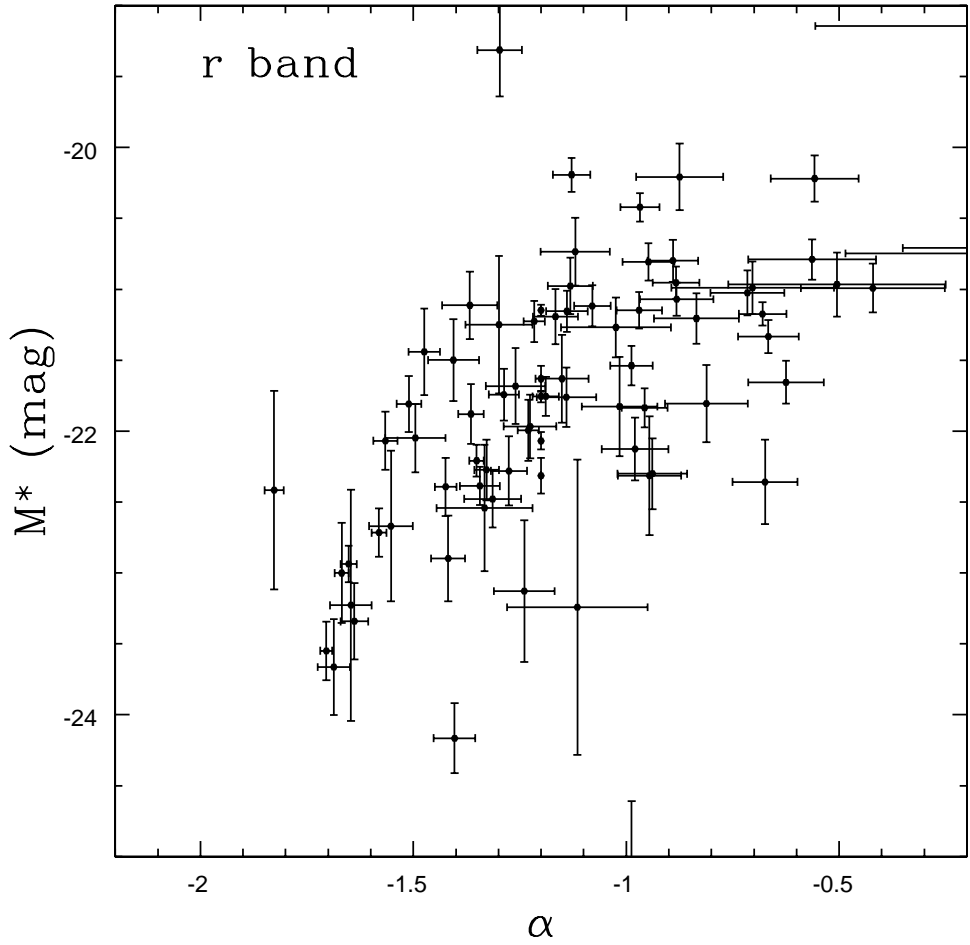}}
\end{minipage}
\end{center}
\caption{
Fit parameters of the individual Schechter luminosity functions in the
r   band.   The fitting  procedure is   performed   by using the local
background magnitude  number counts.    The parameters show  a  slight
correlation.  The   error  bars in  the plot  are  at 68\%  confidence
level.}
\label{mcalpha}
\end{figure}

The  probability that the assumed distribution  gives  a galaxy at the
magnitudes $m_k$ is thus  $\Phi(m_k)$.   Therefore, if the    observed
galaxies are statistically independent,  the combined probability that
the assumed distribution gives the observed  galaxies at the magnitude
$m_k$ (with $k=1,n$) is:
\begin{equation}
L=\prod^n_{\kappa=1}\Phi(m_k)
\end{equation}
The best-fit parameters  are those that maximize  the likelihood L. In
practice   we  have    minimized  the  log-likelihood -2ln(L).    This
minimization was performed with the CERN's software package MINUIT. We
used the variable    metric method MIGRAD    (Flechter 1970) for   the
convergence at the minimum and the MINOS routine to estimate the error
parameters in case of non linearities. We also have placed constraints
on the values   of $m^*$ and $\alpha$  that  the fitting routine   can
accept, to avoid being trapped in  a false minimum  ($M*$ in the range
between -18  and -26 mag and  $\alpha$ between 0  and -2.5, Lumdsen et
al. 1997). Fig. \ref{lf}  gives an example of the  LF derived with MLM
in  each   of       the     Sloan    photometric   bands  for        a
cluster. Figures   \ref{alpha}  and   \ref{mc}  show  the comparison
between the   fit  parameters  calculated with  different  backgrounds
$b(m)$ (local  and  global). Fig. \ref{mcalpha}  shows the correlation
between the fit parameters $M^*$ and $\alpha$.

\begin{figure}
\begin{center}
\begin{minipage}{1.0\textwidth}
\resizebox{\hsize}{!}{\includegraphics{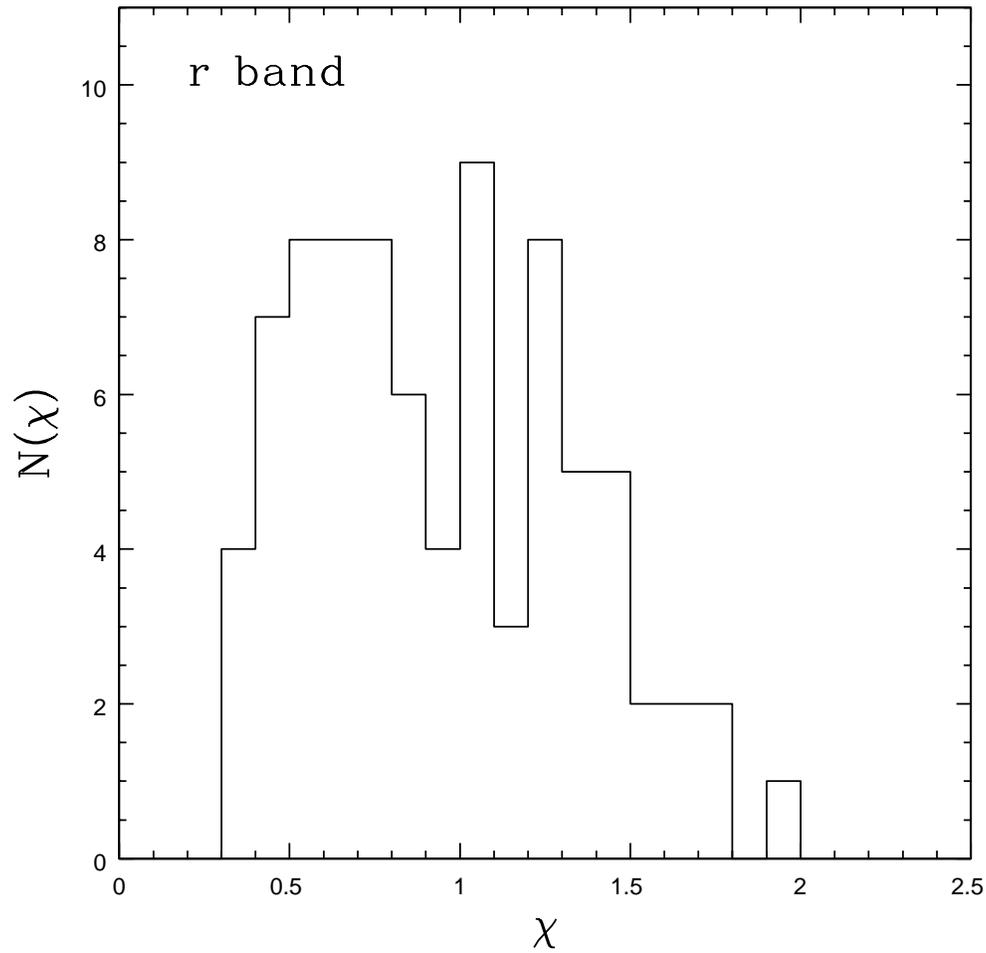}}
\end{minipage}
\end{center}
\caption{
Distribution of  the reduced  $\chi  ^2$  of the individual  Schechter
luminosity  functions.    Since the  MLM allows  us   to perform a fit
without binning   the data but   does not  give information  about the
goodness of the fit, we have performed  the fitting procedure with MLM
and checked  the goodness  of our fitted  luminosity  functions with a
$\chi ^2$ test. All the cluster are well represented by the individual
fitted luminosity function.}
\label{chi}
\end{figure}

The great advantage of the MLM is that the method does not require the
data to be binned  and does not  depend on the  bin size, but uses all
the available  information .  On  the other hand,  the MLM provides no
information  about  the  goodness of the    fit.  Therefore,  we  have
performed  a statistical test.   Since  the routine  procedure uses  a
unbinned set of data to  perform the fit,  in principle the Kolmogorov
Smirnov test should  be applicable to  our case. Nevertheless since the
KS probability is not  easy to interpret, we have  applied a $\chi ^2$
test, by comparing  the background subtracted magnitude number  counts
of the cluster with  the Schechter luminosity  function fitted to  the
data.   Figure \ref{chi} shows the  distribution  of the reduced $\chi
^2$ in the cluster  sample.  Almost 90\% of  the fitted LFs are a good
fit to the data having a reduced $\chi ^2$ $\le 1$.

\subsection{The total luminosity}
\begin{figure}
\begin{center}
\begin{minipage}{1.0\textwidth}
\resizebox{\hsize}{!}{\includegraphics{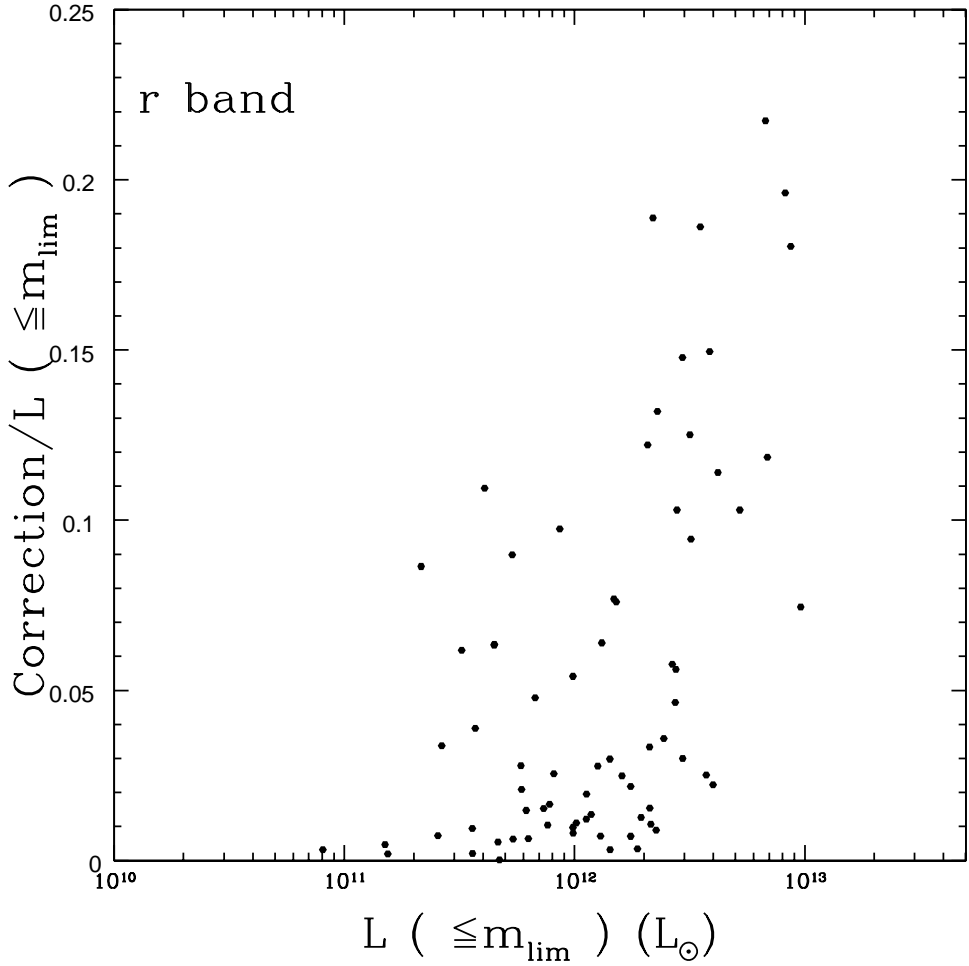}}
\end{minipage}
\end{center}
\caption{
The plot  shows the correction for  incompleteness to the total optical
luminosity.  The completeness magnitude limit  is 21 mag in each Sloan
photometric band.  The correction is  of the order of  5\% for 50\% of
the cluster in the sample, while it is less than 10\%  for 85\% of the
sample. The trend in the plot is due to a  selection effect, since the
most distant cluster are also the most luminous ones.}
\label{incomplete}
\end{figure}

\begin{figure}
\begin{center}
\begin{minipage}{1.0\textwidth}
\resizebox{\hsize}{!}{\includegraphics{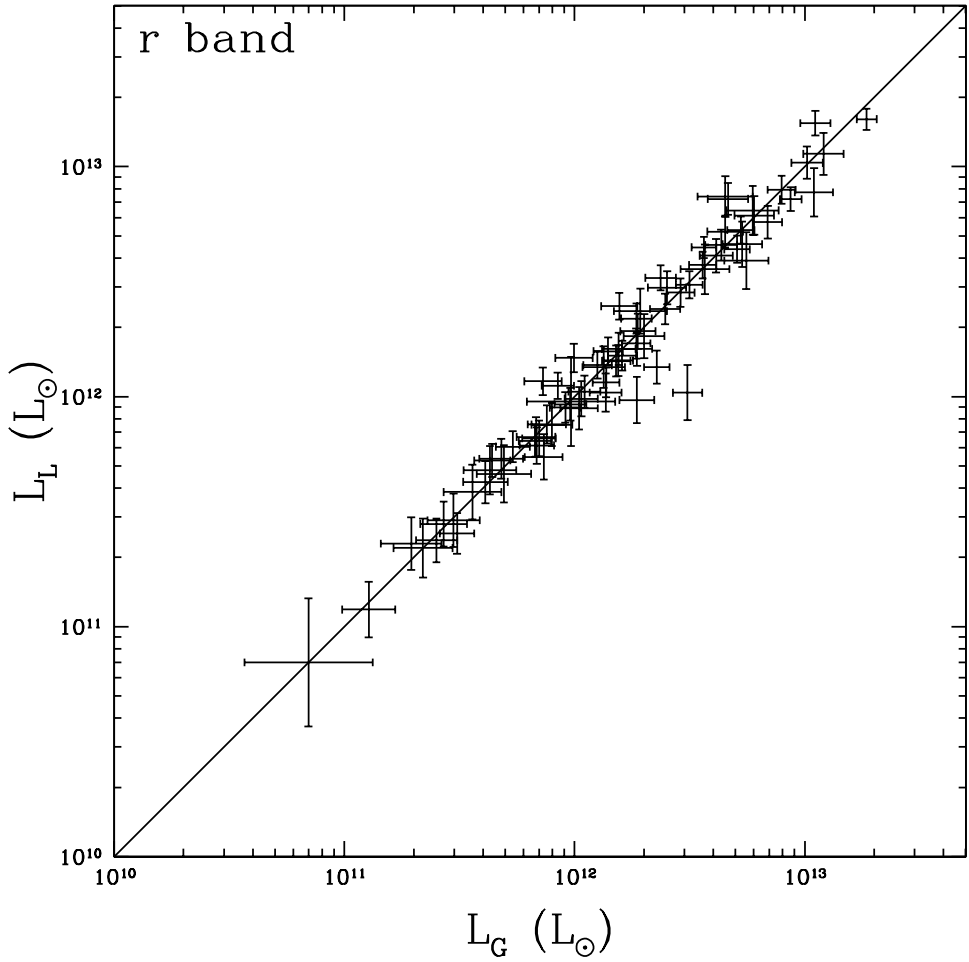}}
\end{minipage}
\end{center}
\caption{
Comparison  of the  cluster  optical luminosities  calculated from the
cluster magnitude number counts with different background corrections.
$L_L$ is calculated with the local background subtraction, while $L_G$
with  the  global one. The   different corrections  do not  affect the
cluster luminosity estimation.  The error bars in the plot are at 68\%
confidence level.}
\label{lm}
\end{figure}

In order to calculate the total cluster luminosity, we have calculated
first the absolute magnitude
\begin{equation}
M=m - 25 -5log_{10}(D_L/1Mpc) - A - K(z)
\end{equation}
where $D_L$ is the  luminosity distance, A  is the Galactic extinction
and $K(z)$ is the  K-correction.  We deredden  the Petrosian and model
magnitudes of  galaxies  using the  Galactic map   of Schlegel et  al.
(1998) in each photometric band.  We used the K-correction supplied by
Fukugita, Shimasaku,  $\&$  Ichikawa (1995)   for elliptical galaxies,
assuming   that  the  main  population of  our    clusters are the old
elliptical galaxies at  the cluster redshift.  The transformation  from
absolute  magnitudes to absolute      luminosity in units   of   solar
luminosities   is performed  by   using the  solar absolute  magnitude
obtained  from     the   color transformation   equation     from  the
Johnson-Morgan-Cousins  system  to  the  SDSS   system of Fukugita  et
al. (1996).
We have  calculated the  optical luminosity  of each cluster  with two
different  methods.    First, we have  estimated    $L$  by using  the
(background corrected) magnitude number counts of the cluster galaxies
with the following prescription:
\begin{equation}
L=\sum_{i}^{N}{N_i(m)l_i(m)}+\int_{m_{lim}}^{\infty}{\phi (m)dm}
\end{equation}
The sum on right  side is performed over  all  the $N$  magnitude bins
with galaxy number $N_i(m)$ and mean luminosity $l_i(m)$. The integral
is an incompleteness  correction due to the  completeness limit of the
galaxy sample at $m_{lim}=21$ mag in the five Sloan photometric bands.
$\phi  (m)$ is the individual  Schechter luminosity function fitted to
the galaxy sample of each cluster. The incompleteness correction is of
the order of 5-10\% in the whole cluster sample, as showed in the Fig.
\ref{incomplete}.  This means   that the galaxies below the  magnitude
limit  do not  give a significant   contribution to  the total optical
luminosity.  Therefore  the most important  source of  error is due to
the contribution   of   the background  galaxy   number  counts.   The
uncertainty in each bin of magnitude is given  by the Poissonian error
of the  bin counts  ($\sqrt{N^i_{tot}(m)}$, with $i=1,...,N$)  and the
background subtraction  in  each magnitude  bin ($\sigma_{bg}^{i}(m)$,
with $i=1,...,N$, see previous section for  the definition). Since the
galaxy counts in the bins are independent, the error in the luminosity
is given by:
\begin{equation}
\Delta L=(\sum_{i}^{N}{(N^i_{tot}(m)+ \sigma_{bg}^{i}(m)^2)})^{\frac{1}{2}}.
\end{equation}
Fig. \ref{lm} shows the  comparison between the luminosity  calculated
from the local background  corrected  and global background  corrected
magnitude  number counts. The difference between  both methods is much
smaller than the statistical error.

In the  second case we have taken  advantage  of the individual fitted
luminosity  function.   If MIGRAD has    converged successfully to the
minimum, we calculate the total luminosity as
\begin{equation}
L=L^*N_{clu}\phi^*\Gamma(2+\alpha)
\label{L}
\end{equation}
where  $L^*$ and $\alpha$  are the  fit parameters  estimated from the
data and $N_{clu}$ is the total number of cluster galaxies.  There are
different sources of errors in this  calculation.  The major source of
error  comes from the  background   which affects  both  the number  of
cluster galaxies $N_{clu}$ and the  result of the fitting procedure. A
second kind of   error   is due  to    the uncertainty of    the  fit
parameters. Since  all these  errors are  not independent, we  can not
treat their contributions  separately. Therefore the  luminosity errors
were  calculated by  varying   the fit  parameter  values,  $M^*$  and
$\alpha$, along  their 68\% confidence  level  error ellipse and using
the upper and  lower  bound of   the quoted  background number  counts
($b(m)$) ranges.  The statistical   luminosity  error range  was  then
defined between the  minimum and maximum  luminosity. With this method
we can take into account statistical and systematic  errors due to the
background and their effects on the  fit parameters as well. Note that
a simple  error  propagation applied to   the equation (\ref{L}) would
underestimate  the  error in the luminosity,  since  it would not take
into account the error of the galaxy background.

\begin{figure}
\begin{center}
\begin{minipage}{1.0\textwidth}
\resizebox{\hsize}{!}{\includegraphics{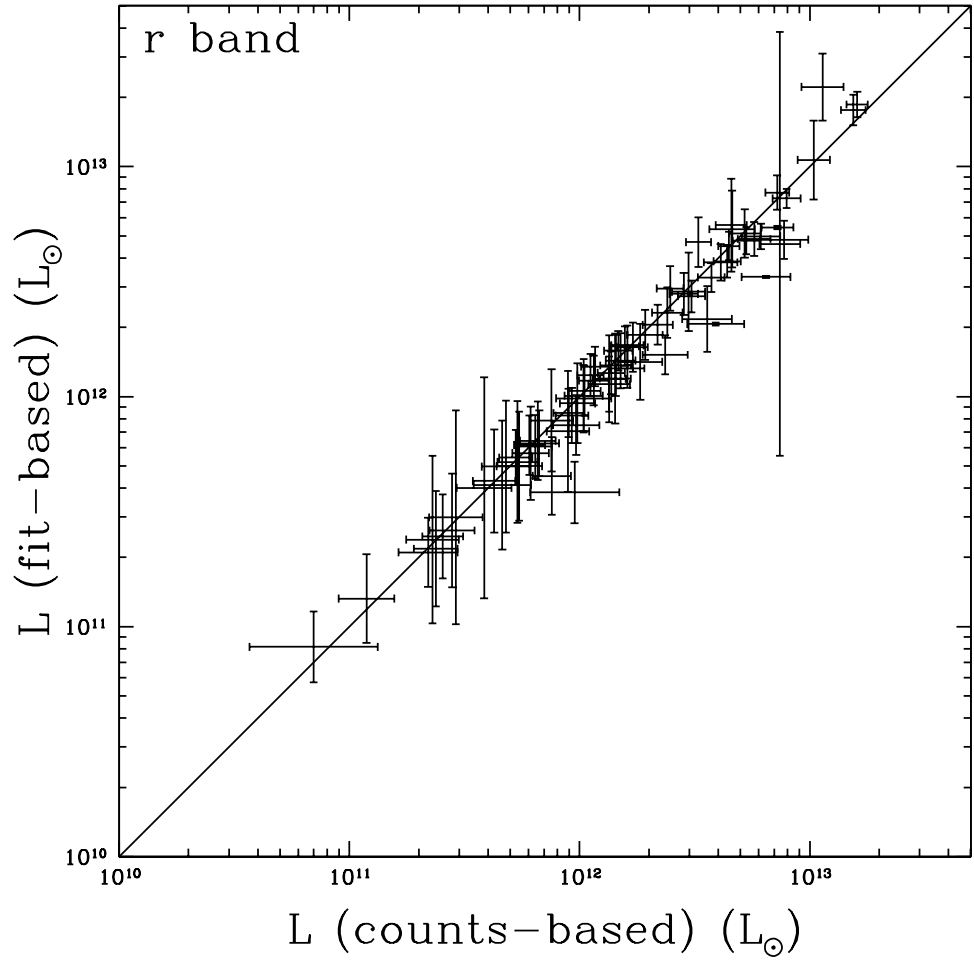}}
\end{minipage}
\end{center}
\caption{
Comparison of the cluster luminosities determined with count-based and
fit-based method.  The error bars  in the plot  are at 68\% confidence
level.}
\label{fc}
\end{figure}

\begin{figure}
\begin{center}
\begin{minipage}{1.0\textwidth}
\resizebox{\hsize}{!}{\includegraphics{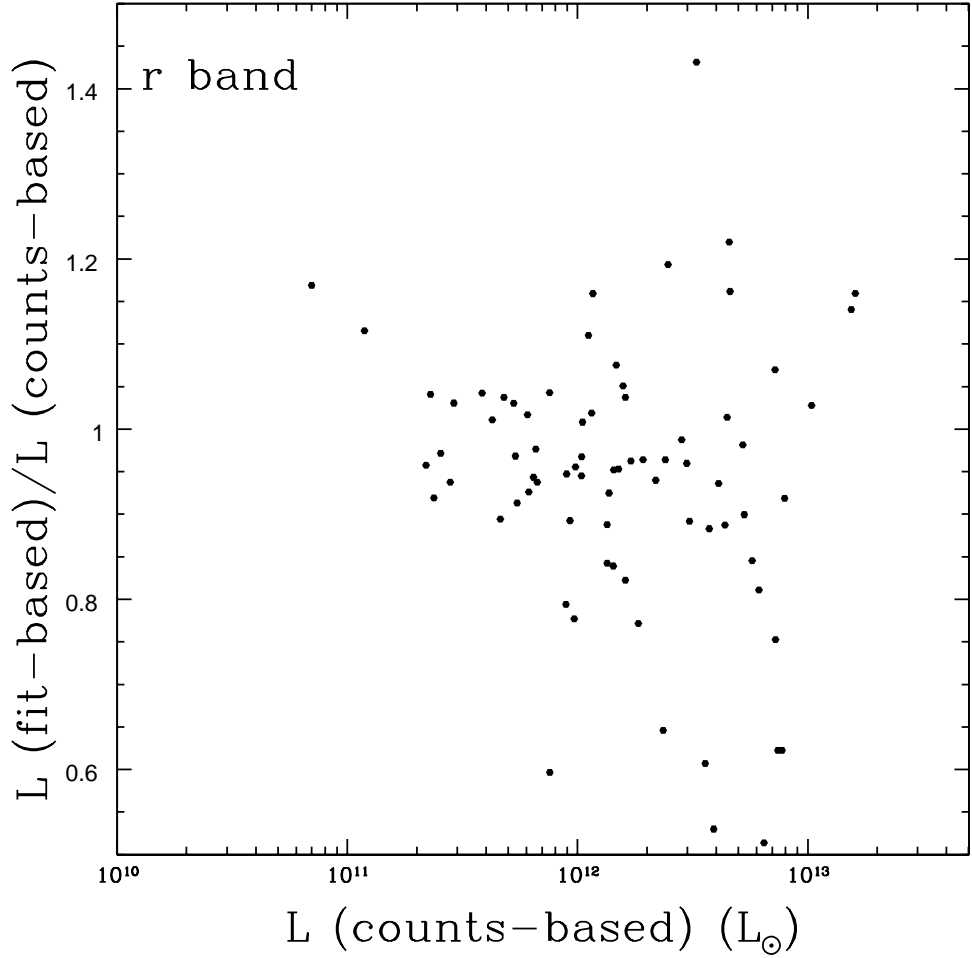}}
\end{minipage}
\end{center}
\caption{
Comparison of  the  cluster  luminosities count-based and  fit-based.
The plot shows  the ratio of  the  two luminosities (count-based  and
fit-based)  versus  the count-based one.  For  70\%  of the sample the
ratio  $L_{fit-based}/L_{count-based}$  is less the  1.1,  indicating
that   fit-based  luminosity   is   systematically  less  bright   the
count-based  one. The reason  is due  to the  Bright Cluster Galaxies
(BCG):  they   are included  in the  computation  of  the count-based
luminosity but not in the estimation of the fit-based one. }
\label{fc2}
\end{figure}

Fig. \ref{fc}  shows  the comparison of  the  two optical luminosities
(fit-based and count-based),  which are consistent within the errors.
For 70\% of the clusters in the sample  the count-based luminosity is
systematically  brighter than the the  fit-based one, as showed in the
Fig. \ref{fc2}. In the former case, indeed, the method includes in the
calculation of $L_{op}$   the Bright Cluster Galaxies  (BCG) which
are usually excluded by the Schechter  luminosity function.  The error
bars in the fit-based luminosity  are larger than in the  count-based
ones. In fact in the former case there are two  main sources of error:
the  uncertainty due  to  the  galaxy  background subtraction  and the
statistical errors in the  fit parameters of the  luminosity function.
   In the latter   case, Instead,  only  the subtraction  of  the galactic
background  plays a  crucial role.  The   mean error in the  fit-based
luminosity  is  around  30\%,  while   it  is   around  20\% with  the
count-based method.

The great advantages of the  count-based optical luminosity is that it
can  be easily computed,  if the cluster $S/N$ is  high enough. On the
other  hand, the fit-based  luminosity  depends on the  success of the
fitting procedure.  Therefore, it is  sensitive not only to the  $S/N$
but also to the    chosen model and the   goodness  of the fit.    The
uncertainty in the count-based method is smaller than in the fit-based
method.  Moreover,  while the count-based  method provides the optical
luminosity  for any system and at  any cluster aperture, the number of
failures in the fitting  procedure is  an  increasing function of  the
cluster aperture.  In fact the fit-based method  fails to fit the data
for 15\% of the clusters at 0.35 Mpc $\rm{h}_{70}^{-1}$ up to 35\%
at 2.0  Mpc  $\rm{h}_{70}^{-1}$.    In consequence,   the count-based
$L_{op}$ has to be preferred to the fit-based  one in the study of the
correlation  between  optical and X-ray  properties.   The count-based
method also reflects what we actually observe.

On the basis  of this analysis we can  conclude that the behavior  of
the    optical  luminosities  calculated   with different   background
subtraction  is stable for variant approaches  and  the main source of
errors   is  due    to    the   necessary    background    subtraction
(Fig.  \ref{lm}). Moreover,   since     the two    different   methods
(count-based and fit-based)  give  consistent results, our  measure of
$L_{op}$ seems to  be a good estimation  of the cluster  total optical
luminosity.

\subsection{The optical structure parameters}
\begin{figure}
\begin{center}
\begin{minipage}{1.0\textwidth}
\resizebox{\hsize}{!}{\includegraphics{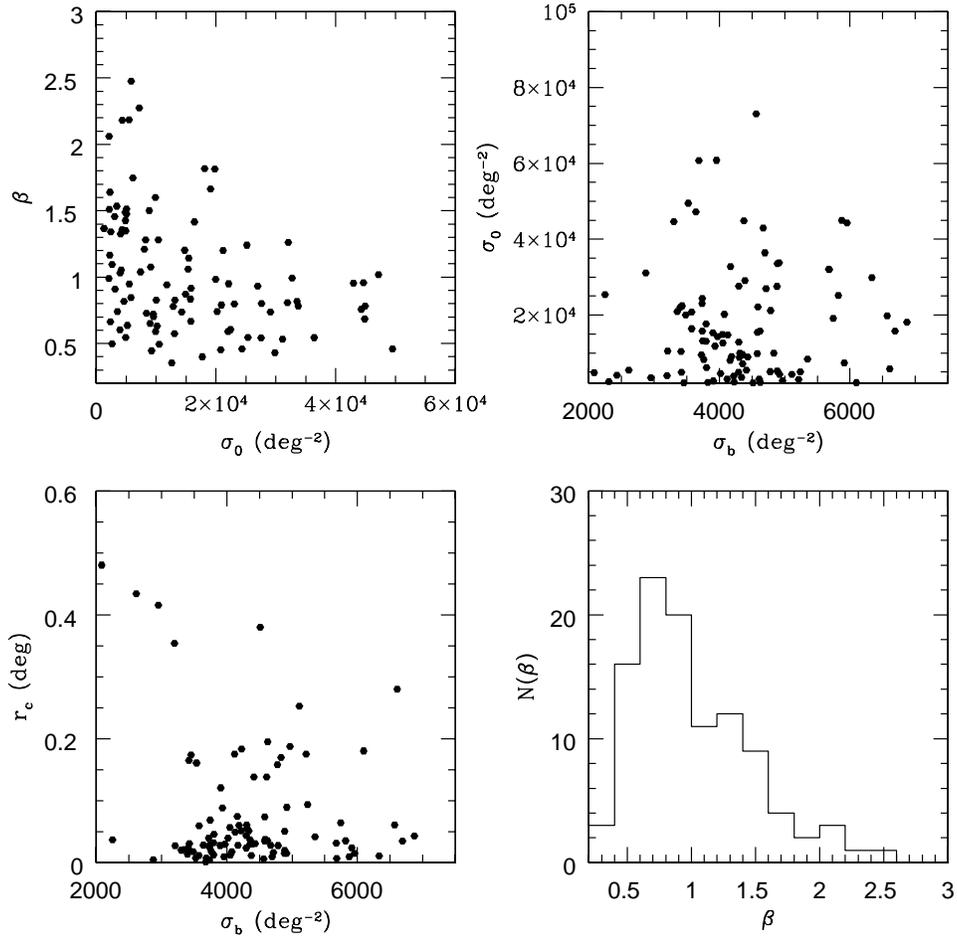}}
\end{minipage}
\end{center}
\caption{
The four panels show the behavior of the fit parameters in the King
profile fit  procedure. The first  three panels show  that there is no
correlation between the    parameters, while the bottom   right  panel
presents the histogram of the exponent $\beta$ of the King profile.}
\label{prof}
\end{figure}

In order to study the spatial distribution of  galaxies in cluster, we
have   analysed  the  projected  radial galaxy   distributions of each
cluster  in the sample.  The analysis is performed in  the g,r i and z
bands.   As for  the luminosity  functions,  we  have  used a  Maximum
Likelihood method to fit a King profile to the data
\begin{equation}
P(r)=\frac{\sigma _0}{(1+(\frac{r}{r_c})^2)^{\beta}}+ \sigma _b.
\label{king}
\end{equation}
In  equation \ref{king}, $\sigma _0 $ is the  central galaxy density, $r_c$ the
core  radius,   $\beta$ the profile   exponent,  and  $\sigma  _b$ the
background density. $P(r)$ has to be normalized through:
\begin{equation}
\int_A{P(r)d(\pi r^2)}=N_{tot}
\label{norm}
\end{equation}
where A is the relevant cluster area and $N_{tot}$ is the total number
of galaxies  within that area. In  agreement with  the section 5.2 the
the Likelihood is given by:
\begin{equation}
L=\prod^n_{\kappa=1} P(r_{\kappa})
\end{equation}
where  $r_{\kappa}$ is the  projected galaxy  distance  from the X-ray
center.  We  regarded  $\beta$,  $r_c$ and   $\sigma _b$  as   fitting
parameters, while $\sigma  _0 $ is a  dependent variable and its value
is derived from the likelihood normalization,  eq.  (\ref{norm}).  The
fitting method worked  successfully in average  for 95\% of the sample
in any photometric  band; it failed for  groups, where the overdensity
in comparison to  the background density is  too low to fit a profile.
As shown in  the fig. \ref{prof},  there are no correlations  between
the parameters, $\sigma _0$,  $r_c$ and  $\beta$, with the  background
density $\sigma _b$.  Furthermore the histogram  of the $\beta$ values
in the same figure  shows that the mean value  of the profile exponent
is around 0.8 with a very large dispersion of 0.5 around the peak.

\begin{figure}
\begin{center}
\begin{minipage}{1.0\textwidth}
\resizebox{\hsize}{!}{\includegraphics{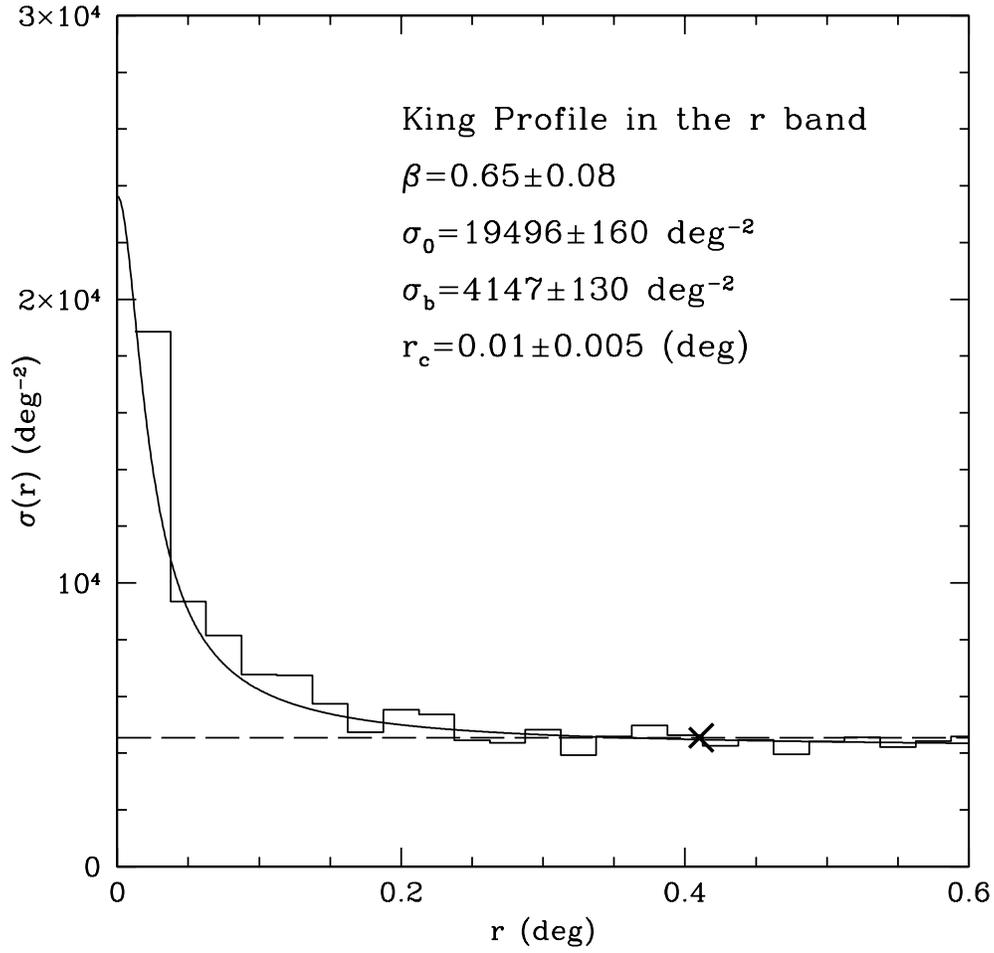}}
\end{minipage}
\end{center}
\caption{
The physical total  size (the cross)  of each cluster is  estimated as
the radius where the galaxy number  density within the cluster becomes
3 times the error in the  statistical background galaxy number density
(the dashed line).}
\label{prof1}
\end{figure}

We have  estimated from the   King profile the  physical size  of each
cluster, $r_{tot}$. We have assumed that this quantity is given by the
radial distance from the X-ray center, where the galaxy number density
of the cluster  becomes $n$ times  the error of  the background galaxy
density ( the  cross in the Fig. \ref{prof1}).   To search for the best
value of $n$, we have estimated the total radius with different values
of $n$  ($n=1,...,5$) and   calculated the  total   optical luminosity
within that radius; $n$ was then fixed to  3, since the differences in
the luminosities calculated  within different total radii are  smaller
than the luminosity uncertainties due to background subtraction.

We assumed that the cluster total optical luminosity in  a band is the
luminosity calculated within $r_{tot}$  estimated in the given filter.
To calculate then the half-light radius ,  which encircles half of the
total   cluster luminosity,  we have     estimated in  each band   the
luminosity of the cluster within 20 radii from the X-ray center to the
total  radius.    We have then  interpolated   in  each filter  the 20
luminosities,  to find  the radius  which  corresponds to half  of the
total cluster luminosity.

\subsection{The catalog}
In the following we  present the catalog of  the 114  RASS-SDSS galaxy
clusters. Tables 1-3 list all the  X-ray and optical properties of the
sample  computed as explained in the   previous sections. The examples
tables show the results  for the first 35 clusters  in the sample. The
complete tables are given in electronic form.

Table (\ref{XX}) give the X-ray properties of the cluster derived from
the ROSAT data. Col. (1) and (2) contain the ROSAT and the alternative
cluster name  respectively.  Col.  (3)  and (4) contain the equatorial
coordinates   of  the  X-ray cluster  center   used   for the regional
selection for  the epoch J2000  in decimal degrees. Col.  (5) contains
the heliocentric cluster  redshift. Col. (6)  presents the flux in the
energy   band   range 0.1-2.4   keV  in   units  of   $10^{-11}$  ergs
$\rm{s}^{-1}$ $\rm{cm}^{-2}$. Col. (7) gives the  corrected flux for a
temperature derived  from $L_X-T$  relation including the K-correction
for an assumed  cluster temperature of  5 keV.  Col.  (8) contains the
relative 1 $\sigma$  Poissonian error of the count  rate, the flux and
the luminosity in percent.  Col (9) gives the luminosity in the energy
range     0.1-2.4 keV   in    units    of $\rm{h}_{70}^{-2}10^{44}$erg
$\rm{s}^{-1}$.  Col.  (10) contains the  count rate in units of counts
$\rm{s}^{-1}$. Col. (11) gives the outer  radius within which the flux
and the luminosity are estimated, in units of arc minutes.

Tables (\ref{LL}) provides  the optical  parameters of the  luminosity
function and the luminosities of each cluster  calculated by using the
local galaxy  background. Listed are results  in the r  band.  All the
quantities are calculated within a fixed  cluster aperture of  1.0
Mpc   $\rm{h}_{70}^{-1}$.  Col.  (1)  presents the  ROSAT name of the
cluster.  Col.   (2) and (3)  show the  resulting fit  parameters of a
Schechter luminosity function.  $\alpha$ is the slope of the LF, while
$M*$  is the magnitude knee of  the distribution.  Col.  (4) gives the
galaxy number density within  the  cluster region selected  to perform
the  fit, in  units of  $\rm{deg}^{-2}$.   Col.  (5) shows the reduced
$\chi ^2$ of  the fitted luminosity   function.  Col (6)  provides the
cluster  optical luminosity   in   units of   $10^{12}$   $L_{\odot}$,
calculated on the basis of  the fitted  luminosity function.  Col  (7)
lists the  cluster    optical   luminosity in    units   of  $10^{12}$
$L_{\odot}$, calculated on  the base of  the cluster magnitude  number
counts.  All the  errors in the table are  at $68\%$ confidence level.
The catalog   contains the extended  version   of this table.  Similar
tables exist, listing the parameters  and the luminosities relative to
22  different cluster apertures:  20 fixed apertures ranging from 
0.05 to 4.0 Mpc $\rm{h}_{70}^{-1}$,  2 variable apertures as the core
radius and the half-light radius. All the data are provided in each of
the 4 Sloan   photometric bands g,r,i  and  z, and for both  local and
global galaxy background correction.

Table  (\ref{RR}) provides the   information concerning to the  radial
distribution  of  the projected  galaxy density in  the  region of the
cluster. Col.  (1) presents the  ROSAT name of  the cluster.  Col. (2)
gives the cluster    central galaxy   number   density  in  units   of
$\rm{deg}^{-2}$  ($\sigma _0$ is  not  a fit  parameter, therefore the
error is not  provided). Col. (3) lists  the background galaxy  number
density  around  the cluster in  units  of $\rm{deg}^{-2}$.   Col. (4)
shows  the   core   radius estimated  from   the   fit, in   units  of
Mpc. Col. (5) provides the cluster total radius, extrapolated from the
King profile, in units of Mpc.  Col. (6) gives the half-light radius
in units of Mpc. All the errors in  the table are at $68\%$ confidence
level.

The full set of extended tables  are available in electronic
form.

\begin{table}
\begin{tabular}[b]{ccccccccccc}\hline\hline
\renewcommand{\arraystretch}{0.2}\renewcommand{\tabcolsep}{0.05cm}
$$  &  Alternative&$$ &$$ &$$ &$$ &$$ &$$ &$$ & Count &$$  \\
Name & Name  & R.A. & dec & $z$ & $F_X$ & ${F_X}^*$ & Error &
 $L_X$ & Rate   & $R_{out}$ \\ 
(1)&(2)&(3)&(4)&(5)&(6)&(7)&(8)&(9)&(10)&(11)\\ \hline

{\bf RXCJ0041.8-0918} & $  A0085       $   &       10.4587  &      -9.3019  &      0.0520   &    67.612   &   67.905   &    3.0   &   7.877  & 3.255  &   18.0 \\
{\bf RXCJ0114.9+0022} & $  A0168       $   &       18.7350  &       0.3746  &      0.0470   &     8.725   &    8.484   &    8.7   &   0.812  & 0.423  &   14.0 \\
{\bf RXCJ0119.6+1453} & $  A0175       $   &       19.9072  &      14.8931  &      0.1290   &     3.124   &    3.114   &   29.1   &   2.237  & 0.148  &    9.5 \\
{\bf RXCJ0137.2-0912} & $  $...$       $  &       24.3140  &      -9.2028  &      0.0390   &     7.275   &    7.071   &    8.4   &   0.464  & 0.358  &    9.5 \\
{\bf RXCJ0152.7+0100} & $  A0267       $   &       28.1762  &       1.0126  &      0.2270   &     4.257   &    4.276   &   12.1   &   9.327  & 0.209  &    9.0 \\
{\bf RXCJ0736.4+3925} & $  $...$       $  &      114.1040  &      39.4329  &      0.1170   &     8.239   &    8.239   &    9.5   &   4.818  & 0.369  &   13.0  \\
{\bf RXCJ0747.0+4131} & $  $...$       $  &      116.7537  &      41.5314  &      0.0280   &     3.201   &    2.431   &   15.0   &   0.083  & 0.147  &    9.5  \\
{\bf RXCJ0753.3+2922} & $  $...$       $  &      118.3291  &      29.3741  &      0.0620   &     6.414   &    0.062   &    9.6   &   1.046  & 0.302  &    9.0 \\
{\bf RXCJ0758.4+3747} & $  Ngc 2484    $   &      119.6172  &      37.7888  &      0.0410   &     0.605   &    0.431   &   32.1   &   0.032  & 0.028  &    7.0 \\
{\bf RXCJ0800.9+3602} & $  A0611       $   &      120.2445  &      36.0469  &      0.2880   &     2.536   &    2.545   &   16.9   &   8.852  & 0.118  &    6.0 \\
{\bf RXCJ0809.6+3455} & $  $...$       $  &      122.4177  &      34.9262  &      0.0800   &     5.208   &    5.164   &   13.2   &   1.436  & 0.242  &    7.5 \\
{\bf RXCJ0810.3+4216} & $  $...$       $  &      122.5942  &      42.2669  &      0.0640   &     2.974   &    2.893   &   13.8   &   0.515  & 0.138  &    6.0  \\
{\bf RXCJ0821.8+0112} & $  $...$       $  &      125.4655  &       1.2116  &      0.0820   &     5.170   &    5.126   &   15.5   &   1.499  & 0.245  &   13.5 \\
{\bf RXCJ0822.1+4705} & $  A0646       $   &      125.5417  &      47.0995  &      0.1300   &     7.236   &    7.236   &    9.0   &   5.253  & 0.346  &    8.0 \\
{\bf RXCJ0824.0+0326} & $  MS0821.5+0337$  &      126.0209  &      3.4383   &      0.3470   &     0.297   &    0.294   &   78.6   &   1.577  & 0.014  &    2.5   \\
{\bf RXCJ0825.4+4707} & $  A0655       $   &      126.3652  &      47.1196  &      0.1260   &     7.235   &    7.235   &   15.1   &   4.926  & 0.342  &   12.0 \\
{\bf RXCJ0828.1+4445} & $  $...$       $  &      127.0278  &      44.7634  &      0.1450   &     4.501   &    4.501   &   11.2   &   4.048  & 0.214  &    6.0 \\
{\bf RXCJ0842.9+3621} & $  A0697       $   &      130.7401  &      36.3625  &      0.2820   &     5.821   &    5.858   &   16.0   &  19.423  & 0.281  &    8.0 \\
{\bf RXCJ0845.3+4430} & $  HGC 35      $   &      131.3434  &      44.5115  &      0.0540   &     0.082   &    0.057   &  100.0   &   0.007  & 0.004  &    0.5  \\
{\bf RXCJ0850.1+3603} & $  CL0847.2+3617$ &      132.5499  &      36.0614  &      0.3730   &     2.742   &    2.75    &   18.7   &  15.876  & 0.134  &    9.5 \\
{\bf RXCJ0913.7+4056} & $  CL09104+4109$   &      138.4411  &      40.9339  &      0.4420   &     1.756   &    1.769   &   30.1   &  14.168  & 0.093  &    8.5   \\
{\bf RXCJ0913.7+4742} & $  A0757       $   &      138.4446  &      47.7021  &      0.0510   &     6.202   &    6.022   &   13.3   &   0.680  & 0.315  &   15.0 \\
{\bf RXCJ0917.8+5143} & $  A0773       $   &      139.4637  &      51.7223  &      0.2170   &     5.961   &    5.998   &    9.2   &  11.853  & 0.305  &    9.0 \\
{\bf RXCJ0943.0+4700} & $  A0851       $   &      145.7600  &      47.0038  &      0.4060   &     1.014   &    1.017   &   30.8   &   7.100  & 0.052  &    4.5  \\
{\bf RXCJ0947.1+5428} & $  $...$       $  &      146.7862  &      54.4754  &      0.0460   &     5.241   &   15.090   &    2.6   &   0.466  & 0.270  &   14.0 \\
{\bf RXCJ0952.8+5153} & $  $...$       $  &      148.2009  &      51.8888  &      0.2140   &     4.196   &    4.216   &   10.6   &   8.131  & 0.218  &    8.0 \\
{\bf RXCJ0953.6+0142} & $  $...$       $  &      148.4231  &       1.7118  &      0.0980   &     2.389   &    2.322   &   24.3   &   0.975  & 0.115  &    9.5  \\
{\bf RXCJ1000.5+4409} & $  $...$       $  &      150.1260  &      44.1550  &      0.1540   &     2.775   &    2.766   &   12.6   &   2.835  & 0.143  &    5.0  \\
{\bf RXCJ1013.7-0006} & $  $...$       $  &      153.4368  &      -0.1085  &      0.0930   &     3.382   &    3.353   &   28.4   &   1.251  & 0.162  &    8.0 \\
{\bf RXCJ1017.5+5933} & $  A0959       $   &      154.3960  &      59.5577  &      0.3530   &     4.079   &    4.109   &   11.4   &  21.151  & 0.211  &   13.0 \\
{\bf RXCJ1022.5+5006} & $  $...$       $  &      155.6283  &      50.1030  &      0.1580   &     5.389   &    5.403   &    9.0   &   5.773  & 0.278  &    8.0  \\
{\bf RXCJ1023.6+0411} & $  $...$       $  &      155.9125  &       4.1873  &      0.2850   &     8.562   &    8.617   &    8.1   &  29.215  & 0.420  &    7.5  \\
{\bf RXCJ1023.6+4908} & $  A0990       $   &      155.9212  &      49.1349  &      0.1440   &     8.180   &    8.202   &    7.3   &   7.236  & 0.422  &   10.0 \\
{\bf RXCJ1053.7+5452} & $  $...$       $  &      163.4349  &      54.8726  &      0.0750   &     4.024   &    3.907   &   11.5   &   0.961  & 0.209  &   11.0  \\
{\bf RXCJ1058.4+5647} & $  $...$       $  &      164.6097  &      56.7922  &      0.1360   &     7.661   &    7.682   &    7.0   &   6.094  & 0.400  &    8.5  \\
\hline
\end{tabular}
\footnotesize
\caption{Example of the table containing the X-ray cluster properties of 
the whole cluster sample. The table show the results for the first 35
cluster in the sample.}
\label{XX}
\end{table}

\normalsize

\begin{table}
\begin{tabular}[b]{ccccccc}\hline\hline
\renewcommand{\arraystretch}{0.2}\renewcommand{\tabcolsep}{0.05cm}
Name & $\alpha$  & $M*$ & $\rho$ & $\chi / \nu$ & $L_F$ & $L_C$ \\ 
\hline
{\bf RXCJ0041.8-0918} & $  -1.20 \pm 0.00$ & $     0.00 \pm 0.00$ &      0 &  0.00 & $   0.000 \pm 0.000$ & $ 1.959 \pm 0.500$\\
{\bf RXCJ0114.9+0022} & $  -1.36 \pm 0.10$ & $   -22.17 \pm 0.70$ &    682 &  1.05 & $   1.049 \pm 0.049$ & $ 1.086 \pm 0.393$\\
{\bf RXCJ0119.6+1453} & $  -1.06 \pm 0.20$ & $   -21.93 \pm 0.69$ &   2775 &  1.21 & $   2.588 \pm 0.071$ & $ 4.128 \pm 1.842$\\
{\bf RXCJ0137.2-0912} & $  -1.20 \pm 0.00$ & $     0.00 \pm 0.00$ &      0 &  0.00 & $   0.000 \pm 0.000$ & $ 0.663 \pm 0.250$\\
{\bf RXCJ0152.7+0100} & $  -1.20 \pm 0.00$ & $     0.00 \pm 0.00$ &      0 &  0.00 & $   0.000 \pm 0.000$ & $ 4.816 \pm 1.719$\\
{\bf RXCJ0736.4+3925} & $  -1.35 \pm 0.19$ & $   -22.30 \pm 1.04$ &   1301 &  0.67 & $   1.775 \pm 0.078$ & $ 1.575 \pm 0.517$\\
{\bf RXCJ0747.0+4131} & $  -1.20 \pm 0.00$ & $     0.00 \pm 0.00$ &      0 &  0.00 & $   0.000 \pm 0.000$ & $ 0.156 \pm 0.103$\\
{\bf RXCJ0753.3+2922} & $  -1.58 \pm 0.09$ & $   -22.78 \pm 0.96$ &   1701 &  1.53 & $   1.103 \pm 0.030$ & $ 1.108 \pm 0.444$\\
{\bf RXCJ0758.4+3747} & $  -1.20 \pm 0.00$ & $     0.00 \pm 0.00$ &      0 &  0.00 & $   0.000 \pm 0.000$ & $ 0.284 \pm 0.190$\\
{\bf RXCJ0800.9+3602} & $  -0.75 \pm 1.06$ & $   -21.15 \pm 1.45$ &   5733 &  1.28 & $   2.429 \pm 0.062$ & $ 4.796 \pm 3.073$\\
{\bf RXCJ0809.6+3455} & $  -1.35 \pm 0.12$ & $   -21.50 \pm 0.63$ &   1850 &  0.56 & $   1.266 \pm 0.023$ & $ 1.247 \pm 0.372$\\
{\bf RXCJ0810.3+4216} & $  -1.56 \pm 0.14$ & $   -22.33 \pm 1.27$ &    975 &  0.57 & $   0.593 \pm 0.081$ & $ 0.611 \pm 0.272$\\
{\bf RXCJ0821.8+0112} & $  -0.71 \pm 0.30$ & $   -21.42 \pm 0.55$ &   1234 &  0.68 & $   1.507 \pm 0.026$ & $ 1.577 \pm 0.548$\\
{\bf RXCJ0822.1+4705} & $  -1.20 \pm 0.00$ & $     0.00 \pm 0.00$ &      0 &  0.00 & $   0.000 \pm 0.000$ & $ 0.913 \pm 0.455$\\
{\bf RXCJ0824.0+0326} & $  -0.77 \pm 0.20$ & $   -20.60 \pm 0.37$ &   3250 &  0.56 & $   2.185 \pm 0.027$ & $ 2.429 \pm 0.677$\\
{\bf RXCJ0825.4+4707} & $  -1.61 \pm 0.19$ & $   -23.12 \pm 1.49$ &   2641 &  0.65 & $   1.743 \pm 0.099$ & $ 1.470 \pm 0.559$\\
{\bf RXCJ0828.1+4445} & $  -1.41 \pm 0.64$ & $   -21.33 \pm 1.18$ &   6881 &  0.28 & $   3.573 \pm 0.148$ & $ 2.385 \pm 0.746$\\
{\bf RXCJ0842.9+3621} & $   0.00 \pm 0.57$ & $   -19.97 \pm 0.75$ &     20 &  1.06 & $   0.182 \pm 0.022$ & $ 0.238 \pm 0.145$\\
{\bf RXCJ0845.3+4430} & $  -1.20 \pm 0.00$ & $     0.00 \pm 0.00$ &      0 &  0.00 & $   0.000 \pm 0.000$ & $ 3.972 \pm 1.351$\\
{\bf RXCJ0850.1+3603} & $  -1.20 \pm 0.00$ & $     0.00 \pm 0.00$ &      0 &  0.00 & $   0.000 \pm 0.000$ & $ 1.655 \pm 1.364$\\
{\bf RXCJ0913.7+4056} & $  -1.24 \pm 0.20$ & $   -21.31 \pm 0.89$ &    308 &  0.68 & $   0.474 \pm 0.017$ & $ 0.476 \pm 0.194$\\
{\bf RXCJ0913.7+4742} & $  -1.06 \pm 0.31$ & $   -21.35 \pm 0.63$ &   7365 &  0.72 & $   3.606 \pm 0.075$ & $ 3.795 \pm 1.064$\\
{\bf RXCJ0917.8+5143} & $  -1.20 \pm 0.00$ & $     0.00 \pm 0.00$ &      0 &  0.00 & $   0.000 \pm 0.000$ & $ 5.492 \pm 1.971$\\
{\bf RXCJ0943.0+4700} & $  -0.01 \pm 0.51$ & $   -19.67 \pm 0.47$ &     54 &  1.07 & $   0.681 \pm 0.075$ & $ 0.798 \pm 0.244$\\
{\bf RXCJ0947.1+5428} & $   0.00 \pm 0.90$ & $   -20.23 \pm 0.87$ &   1716 &  0.31 & $   0.892 \pm 0.011$ & $ 1.483 \pm 0.696$\\
{\bf RXCJ0952.8+5153} & $  -2.16 \pm 1.45$ & $   -25.53 \pm 0.00$ &    958 &  0.72 & $   0.739 \pm 0.000$ & $ 0.904 \pm 0.865$\\
{\bf RXCJ0953.6+0142} & $  -0.69 \pm 0.89$ & $   -21.24 \pm 1.67$ &    411 &  0.48 & $   0.505 \pm 0.107$ & $ 0.532 \pm 0.285$\\
{\bf RXCJ1000.5+4409} & $  -1.33 \pm 0.27$ & $   -20.91 \pm 0.95$ &   1150 &  0.95 & $   0.597 \pm 0.018$ & $ 0.617 \pm 0.286$\\
{\bf RXCJ1013.7-0006} & $  -1.20 \pm 0.00$ & $     0.00 \pm 0.00$ &      0 &  0.00 & $   0.000 \pm 0.000$ & $ 5.320 \pm 1.670$\\
{\bf RXCJ1017.5+5933} & $  -1.24 \pm 0.23$ & $   -21.66 \pm 0.75$ &   3959 &  0.69 & $   2.417 \pm 0.076$ & $ 2.434 \pm 0.717$\\
{\bf RXCJ1022.5+5006} & $  -0.65 \pm 0.72$ & $   -20.52 \pm 0.65$ &   5544 &  0.21 & $   2.265 \pm 0.032$ & $ 2.348 \pm 0.981$\\
{\bf RXCJ1023.6+0411} & $  -0.75 \pm 0.40$ & $   -20.58 \pm 0.73$ &   2442 &  0.60 & $   1.079 \pm 0.036$ & $ 1.309 \pm 0.492$\\
{\bf RXCJ1023.6+4908} & $  -1.31 \pm 0.13$ & $   -21.80 \pm 0.65$ &   1227 &  1.06 & $   1.016 \pm 0.010$ & $ 1.036 \pm 0.358$\\
{\bf RXCJ1053.7+5452} & $  -1.52 \pm 0.21$ & $   -21.78 \pm 1.21$ &   3964 &  1.25 & $   2.397 \pm 0.203$ & $ 2.318 \pm 0.750$\\
{\bf RXCJ1058.4+5647} & $  -1.49 \pm 0.09$ & $   -21.98 \pm 0.64$ &   2557 &  0.63 & $   1.338 \pm 0.011$ & $ 1.338 \pm 0.409$\\

\hline
\end{tabular}
\caption{Example of the table containing optical properties of the whole 
cluster sample in the r band. The table show the results for the first 35 clusters 
in the sample.}
\label{LL}
\end{table}

\begin{table}
\begin{tabular}[b]{cccccc}\hline\hline
\renewcommand{\arraystretch}{0.2}\renewcommand{\tabcolsep}{0.05cm}
Name & $\sigma _0$ & $\sigma _b$ & $r_c$ & $r_{tot}$ & $r_h$ \\
\hline
{\bf RXCJ0041.8-0918}   &   5510 & $  2614 \pm    204$ & $ 0.434 \pm 0.015$ & 3.362 & $ 1.229 \pm  0.275$ \\
{\bf RXCJ0114.9+0022}   &   2299 & $  4626 \pm     52$ & $ 0.195 \pm 0.026$ & 0.940 & $ 0.365 \pm  0.061$ \\
{\bf RXCJ0119.6+1453}   &  17720 & $  3796 \pm    498$ & $ 0.013 \pm 0.003$ & 1.599 & $ 0.807 \pm  0.068$ \\
{\bf RXCJ0137.2-0912}   &   5071 & $  5239 \pm     27$ & $ 0.094 \pm 0.013$ & 0.697 & $ 0.300 \pm  0.127$ \\
{\bf RXCJ0152.7+0100}   &  12846 & $  4297 \pm    607$ & $ 0.044 \pm 0.007$ & 1.344 & $ 0.866 \pm  0.103$ \\
{\bf RXCJ0736.4+3925}   &   4834 & $  2083 \pm    594$ & $ 0.480 \pm 0.022$ & 2.516 & $ 1.269 \pm  0.188$ \\
{\bf RXCJ0747.0+4131}   &      0 & $     0 \pm      0$ & $ 0.000 \pm 0.000$ & 0.000 & $ 0.000 \pm  0.000$ \\
{\bf RXCJ0753.3+2922}   &   5081 & $  4776 \pm     94$ & $ 0.158 \pm 0.014$ & 1.178 & $ 0.413 \pm  0.053$ \\
{\bf RXCJ0758.4+3747}   &      0 & $     0 \pm      0$ & $ 0.000 \pm 0.000$ & 0.000 & $ 0.000 \pm  0.000$ \\
{\bf RXCJ0800.9+3602}   &  27566 & $  4883 \pm    382$ & $ 0.015 \pm 0.005$ & 3.044 & $ 1.295 \pm  0.581$ \\
{\bf RXCJ0809.6+3455}   &   4168 & $  2428 \pm   2908$ & $ 0.688 \pm 0.000$ & 0.000 & $ 0.000 \pm  0.000$ \\
{\bf RXCJ0810.3+4216}   &   8986 & $  4433 \pm    152$ & $ 0.031 \pm 0.009$ & 0.935 & $ 0.397 \pm  0.034$ \\
{\bf RXCJ0821.8+0112}   &   8840 & $  4306 \pm     41$ & $ 0.061 \pm 0.008$ & 1.422 & $ 0.553 \pm  0.076$ \\
{\bf RXCJ0822.1+4705}   &  31953 & $  5683 \pm    153$ & $ 0.007 \pm 0.003$ & 1.183 & $ 0.425 \pm  0.107$ \\
{\bf RXCJ0824.0+0326}   &  14918 & $  4046 \pm    150$ & $ 0.057 \pm 0.006$ & 2.404 & $ 0.762 \pm  0.223$ \\
{\bf RXCJ0825.4+4707}   &  44865 & $  4375 \pm    157$ & $ 0.011 \pm 0.003$ & 2.022 & $ 0.506 \pm  0.100$ \\
{\bf RXCJ0828.1+4445}   &  49497 & $  3524 \pm    396$ & $ 0.007 \pm 0.003$ & 4.451 & $ 2.087 \pm  0.548$ \\
{\bf RXCJ0842.9+3621}   &   8253 & $  3760 \pm     91$ & $ 0.018 \pm 0.007$ & 0.179 & $ 0.038 \pm  0.047$ \\
{\bf RXCJ0845.3+4430}   &  25163 & $  5821 \pm    106$ & $ 0.035 \pm 0.004$ & 2.633 & $ 1.477 \pm  0.243$ \\
{\bf RXCJ0850.1+3603}   &  31130 & $  2874 \pm    402$ & $ 0.005 \pm 0.003$ & 1.479 & $ 1.317 \pm  0.063$ \\
{\bf RXCJ0913.7+4056}   &   3537 & $  4338 \pm    118$ & $ 0.051 \pm 0.014$ & 0.593 & $ 0.349 \pm  0.050$ \\
{\bf RXCJ0913.7+4742}   &  42986 & $  4670 \pm    116$ & $ 0.028 \pm 0.003$ & 3.090 & $ 0.602 \pm  0.157$ \\
{\bf RXCJ0917.8+5143}   &  23067 & $  3733 \pm    227$ & $ 0.025 \pm 0.005$ & 3.137 & $ 0.884 \pm  0.183$ \\
{\bf RXCJ0943.0+4700}   &      0 & $     0 \pm      0$ & $ 0.000 \pm 0.000$ & 0.000 & $ 0.000 \pm  0.000$ \\
{\bf RXCJ0947.1+5428}   &  15331 & $  3898 \pm    106$ & $ 0.028 \pm 0.006$ & 1.483 & $ 0.579 \pm  0.090$ \\
{\bf RXCJ0952.8+5153}   &   3956 & $  4219 \pm    493$ & $ 0.051 \pm 0.017$ & 0.476 & $ 0.270 \pm  0.034$ \\
{\bf RXCJ0953.6+0142}   &  19978 & $  3484 \pm     85$ & $ 0.017 \pm 0.004$ & 1.065 & $ 0.256 \pm  0.063$ \\
{\bf RXCJ1000.5+4409}   &  15457 & $  4581 \pm     53$ & $ 0.035 \pm 0.006$ & 1.134 & $ 0.215 \pm  0.049$ \\
{\bf RXCJ1013.7-0006}   &  44634 & $  3303 \pm    165$ & $ 0.020 \pm 0.003$ & 2.617 & $ 0.883 \pm  0.174$ \\
{\bf RXCJ1017.5+5933}   &  33538 & $  4887 \pm    121$ & $ 0.020 \pm 0.003$ & 2.166 & $ 0.991 \pm  0.300$ \\
{\bf RXCJ1022.5+5006}   &  73049 & $  4563 \pm    242$ & $ 0.006 \pm 0.003$ & 2.741 & $ 1.516 \pm  0.178$ \\
{\bf RXCJ1023.6+0411}   &  14797 & $  4129 \pm     62$ & $ 0.049 \pm 0.006$ & 1.907 & $ 0.337 \pm  0.036$ \\
{\bf RXCJ1023.6+4908}   &   3003 & $  5215 \pm     53$ & $ 0.175 \pm 0.027$ & 0.712 & $ 0.309 \pm  0.066$ \\
{\bf RXCJ1053.7+5452}   &  22426 & $  3428 \pm    265$ & $ 0.031 \pm 0.005$ & 2.950 & $ 1.064 \pm  0.115$ \\
{\bf RXCJ1058.4+5647}   &   4323 & $  5110 \pm    135$ & $ 0.253 \pm 0.019$ & 1.269 & $ 0.450 \pm  0.088$ \\
			         
\hline			         
\end{tabular}		         
\caption{Example of the table    containing the optical properties of the 
whole cluster sample in the r band. The table shows the results for
the first 35 cluster of the sample. }
\label{RR}
\end{table}

\section{Correlating X-ray and optical properties}

For a  cluster  in which  mass traces   optical light ($M/L_{opt}$  is
constant), the gas is in hyrostatic equilibrium ($T \propto M^{2/3}$),
and $L_X \propto T^3$  (Xue \& Wu 2000), we expect the X-ray
bolometric  luminosity   to be  related  to the  optical luminosity as
$L_{op}
\propto L_{X}^{0.5}$ and to the intracluster medium temperature as 
$L_{op} \propto T_{X}^{1.5}$. 

We have now an optimal data base to test these scaling relations. In a
first step we look for those optical  parameters which are best suited
for  a correlation analysis  and apply these in   a second step to the
test of the scaling relations.

In this section  we show  that tight  correlations  exist between  the
total optical cluster luminosity and  the X-ray cluster properties  as
the X-ray luminosity and the intracluster medium temperature.

To    search for  the   best  correlation between   optical  and X-ray
properties and  to optimally predict for  example the X-ray luminosity
from     the optical appearance,  we   are   interested in an  optical
characteristic, which  shows a  minimum scatter in   the X-ray/optical
correlation. Therefore, we performe a correlation using 4 of the 5 SDSS
optical band, $g$, $r$, $i$  and $z$, to  find out which filter should
be used in the prediction. The $u$ band was not used since the cluster
signal to noise in that band is too low to calculate the cluster total
luminosity.   We  used a  fixed   aperture  to calculate  the  optical
luminosities  for all the clusters,  to   make no a priori  assumption
about the cluster size. Moreover, to check whether  the scatter in the
correlation depends on the cluster aperture, we  did the same analysis
several time by using optical luminosities calculated within different
radii, ranging from  0.05  to 4 Mpc $\rm{h}_{70}^{-1}$  from the
X-ray center.  To quantify the $L_{op} - L_X$ and  the $L_{op} - T_x $
relations, a linear regression in log-log space was performed by using
two    methods   for the  fitting:  a   numerical  orthogonal distance
regression method (ODRPACK)  and   the bisector method (Akritas   $\&$
Bershady 1996).  The fits are performed using the form
\begin{equation}
log(L_{op}/L_{\odot})=\alpha log(P_X)+{\beta}
\end{equation}
where $P_X$ is the X-ray property, and the errors of each variable are
transformed in  to log  space as $\Delta log(x)=log(e)(x^+-x^-)/(2x)$,
where  $x^+$ and  $x^-$ denote the   upper and  lower boundary of  the
quantity's error range, respectively.   To exclude the outliers in the
fitting procedure we apply a  $\sigma$ clipping method.  After a first
fit all  the points featuring a  larger than  $3\sigma$ deviation from
the relation,   were excluded and  the  fitting procedure was repeated
(see discussion below).

\begin{figure}
\begin{center}
\begin{minipage}{1.0\textwidth}
\resizebox{\hsize}{!}{\includegraphics{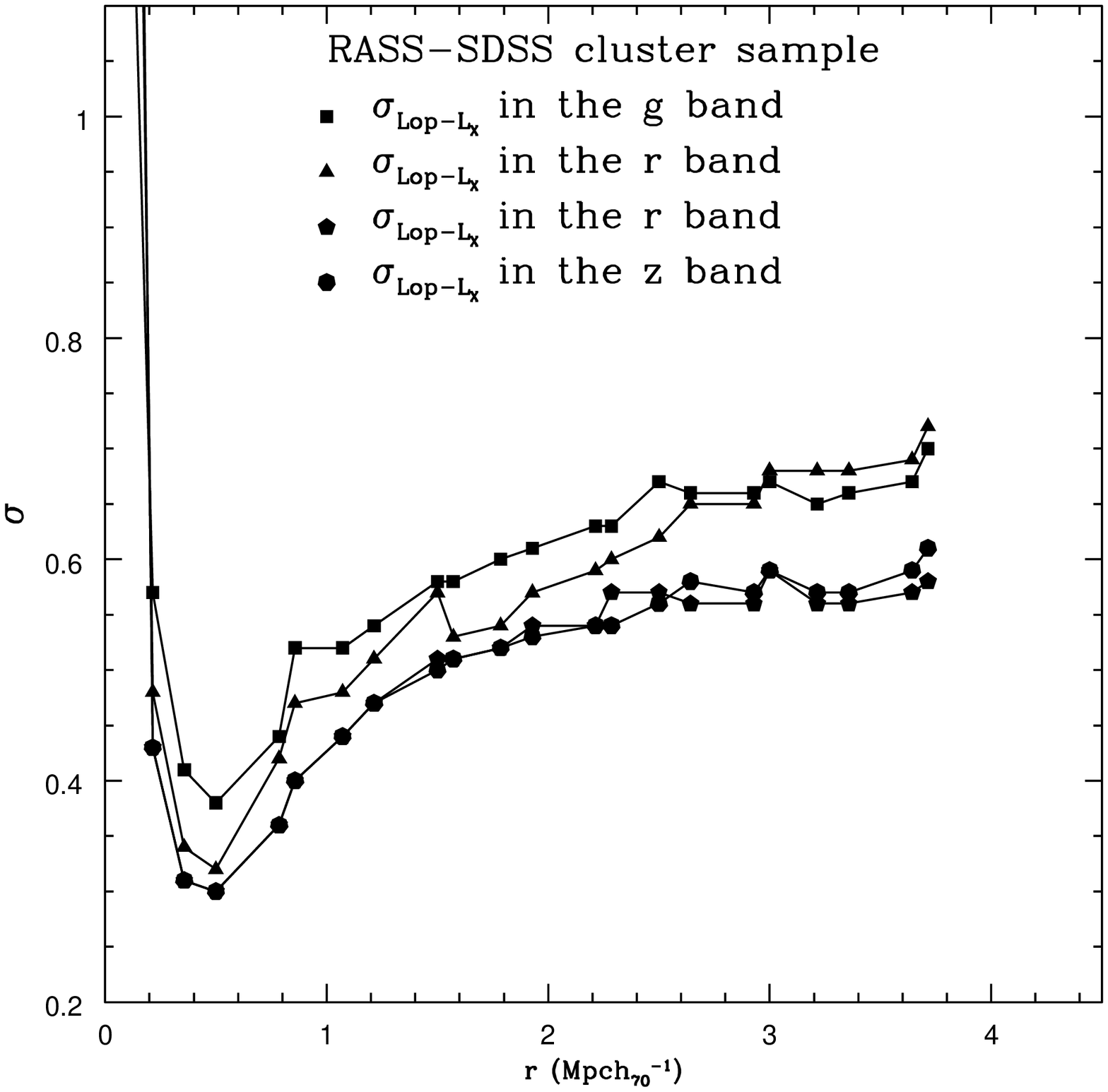}}
\end{minipage}
\end{center}
\caption{
Orthogonal scatter to  the best fit of  the  $L_{opt} - L_X$  relation
obtained with  the  orthogonal method,  as a function  of  the cluster
aperture.  The scatter  shows a minimum in  the region $0.2  \le r \le
0.8 $Mpc $\rm{h}_{70}^{-1}$.    In  the  plot different symbols    are
related  to the different    photometric bands  in  which $L_{op}$   is
calculated:  squares  for the   g   band, triangles for  the  r  band,
hexagons for i and filled  circles for the z  band. The i and z bands
show clearly the smallest scatter at any aperture.}
\label{scatter1}
\end{figure}

\begin{figure}
\begin{center}
\begin{minipage}{1.0\textwidth}
\resizebox{\hsize}{!}{\includegraphics{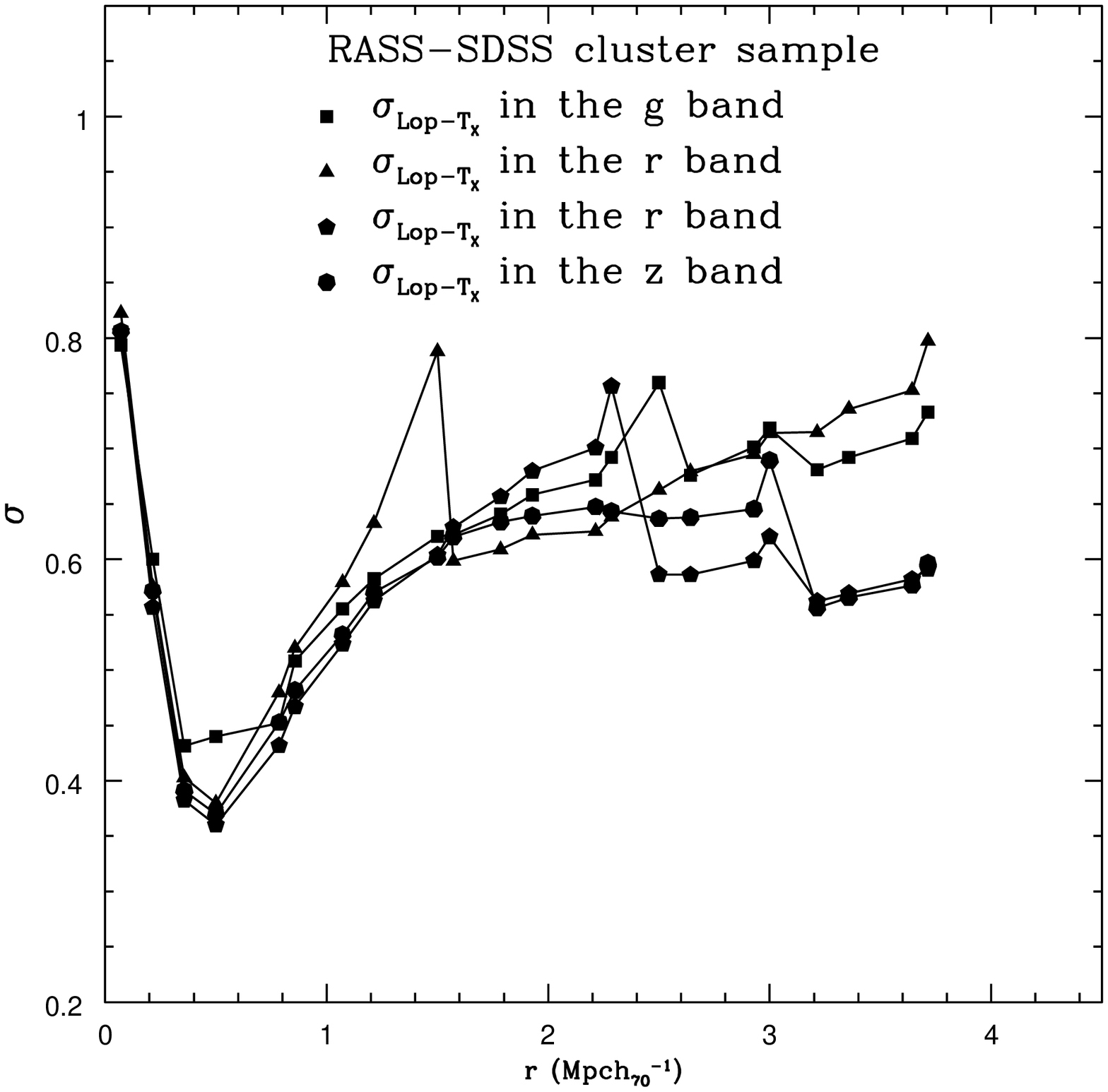}}
\end{minipage}
\end{center}
\caption{
Orthogonal  scatter to  the best fit  of  the $L_{opt} - T_X$ relation
obtained with   the orthogonal method,  as a  function  of the cluster
aperture.  The scatter  shows a minimum in the  region $0.2 \le r  \le
0.8$ Mpc $\rm{h}_{70}^{-1}$.   In   the plot  different symbols  are
related  to  the  different photometric  bands  in  which  $L_{op}$ is
calculated: squares   for  the g   band,  triangles  for the   r band,
hexagons for i and filled  circles for the z  band. The i and z bands
show clearly the smallest scatter at any aperture.}
\label{scatter2}
\end{figure}

Figs.  \ref{scatter1}  and  \ref{scatter2}  show the  scatter   of the
$L_{op} - L_X$ and the $L_{op} - T_x $ relations, respectively, versus
the  cluster aperture, used to calculate  the optical luminosity.  The
scatter in the plot is the orthogonal scatter  estimated from the best
fit given by ODRPACK. In any photometric band the  scatter has a clear
dependence on the  cluster aperture by  showing a region of minimum on
the   very     center  of    the  cluster,  between       0.2  Mpc
$\rm{h}_{70}^{-1}$ and 0.8  Mpc $\rm{h}_{70}^{-1}$, and by increasing
at  larger radii.  The  source of the scatter is  partially due to the
method used to calculate the optical luminosity.  Our method is simply
based on  the overdensity of the  cluster  with respect  of the galaxy
background.  In   fact if the  central   region in  which $L_{op}$  is
measured   is  to small (  0.05  Mpc  $\rm{h}_{70}^{-1}$ in Figs.
\ref{scatter1} and \ref{scatter2}), the value of the galaxy density is
low and the measurement becomes more uncertain.  On the other hand, at
larger radii the    density contrast between  cluster  and  background
decreases progressively.  Instead,  within a  cluster aperture between
 0.2 Mpc  $\rm{h}_{70}^{-1}$ and 0.8 Mpc $\rm{h}_{70}^{-1}$,  the
optical luminosity of both  groups and massive  clusters can be easily
measured.  In  fact in both cases the  radial aperture is small enough
to show an high density contrast and  therefore an high cluster signal
to   noise,   and still large  to  enclose   enough  galaxies  for the
luminosity calculation.

  After a  more accurate  analysis,  we noted  also that  the  low
luminosity systems (both in the optical and in the X-ray band) are the
main source of scatter at any cluster  aperture.  This could be due to
different reasons. From the technical point  of view the groups have a
lower surface   density   contrast,  and   this causes   problems   in
calculating the  optical    luminosity with  a method   based   on the
overdensity contrast.  Moreover,  the  low mass  systems could have  a
larger   scatter in the optical and   X-ray properties.  

\begin{figure}
\begin{center}
\begin{minipage}{1.0\textwidth}
\resizebox{\hsize}{!}{\includegraphics{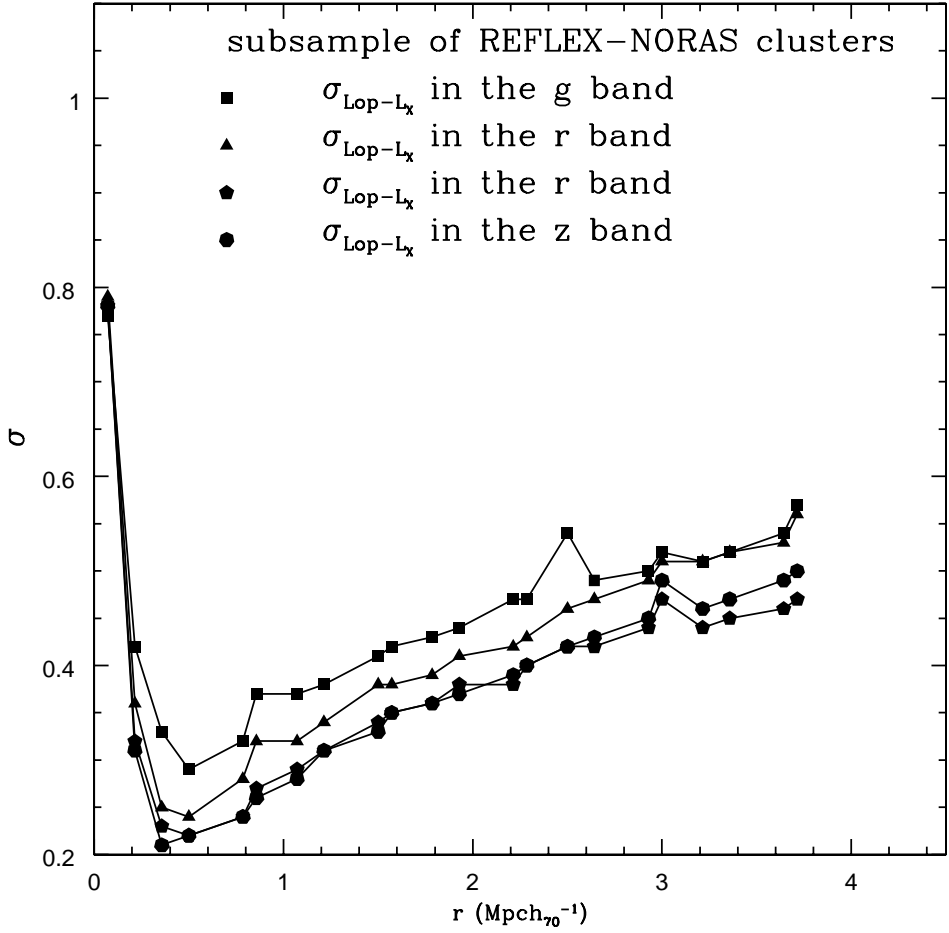}}
\end{minipage}
\end{center}
\caption{
Orthogonal scatter to the  best fit  of  the $L_{opt} -  L_X$ relation
obtained  with the  orthogonal  method, as a  function  of the cluster
aperture  for  the subsample of    $REFLEX-NORAS$ clusters.  The scatter
shows  again  a minimum    in the  region  $0.2 \le    r \le  0.8 $Mpc
$\rm{h}_{70}^{-1}$.  In the plot different symbols  are related to the
different  photometric bands in which  $L_{op}$ is calculated: squares
for the g band,  triangles for the r band,  hexagons for i and  filled
circles for the z  band. The i and  z bands show clearly the  smallest
scatter at any aperture.}
\label{scatter3}
\end{figure}

Furthermore, the galaxy groups could  be responsible for the  behavior
of  the scatter  shown in   fig.   \ref{scatter1}.  In  fact at  large
cluster apertures the galaxy density contrast can  be very low for the
small  systems   and  still  very high   for  the  massive and  larger
clusters. The large error  introduced by the  low density  contrast in
the calculation of  the optical luminosity of  galaxy groups could  be
the source of scatter able to explain the increment  of the scatter at
larger apertures.  In order to study in more details the nature of the
scatter of our correlation and in order to investigate the role of the
less luminous systems, we carried   out the analysis explained   above
with  the low mass systems  removed.  We limited   the analysis to the
subsample of the  X-ray selected  REFLEX-NORAS clusters, which  occupy
the intermediate  and  high luminosity region.    Fig.  \ref{scatter3}
shows   the behavior  of  the scatter  as a   function  of the cluster
aperture in this second analysis.  After removing the low mass systems
the scatter decreases by about 30\% for any  aperture (from 0.3 dex to
0.2 dex in the  region of minimum scatter and  from 0.6 dex in average
to 0.4   dex at larger apertures).   Nevertheless  the behavior of the
scatter at increasing aperture is  absolutely the same observed in the
analysis  carried   out   on the  overall    RASS-SDSS galaxy  cluster
sample. This means that groups are responsible for part of the scatter
but can not  explain the existence  of  the region of  minimum scatter
between 0.2 and  0.8  Mpc $\rm{h}_{70}^{-1}$.  A  possible explanation
for  the behavior of the  scatter at different cluster apertures could
be the cluster compactness.  As $L_X$ depends  not only on the cluster
mass  but also on  the compactness  of the  cluster, also the  optical
luminosity should  reflect somehow the   cluster properties, mass  and
concentration.  Thus, there   should  be an optimal   aperture  radius
within which to  measure $L_{op}$.  We  found that in  all photometric
bands, the minimum scatter is around 0.5 Mpc $\rm{h}_{70}^{-1}$.

\begin{figure}
\begin{center}
\begin{minipage}{1.0\textwidth}
\resizebox{\hsize}{!}{\includegraphics{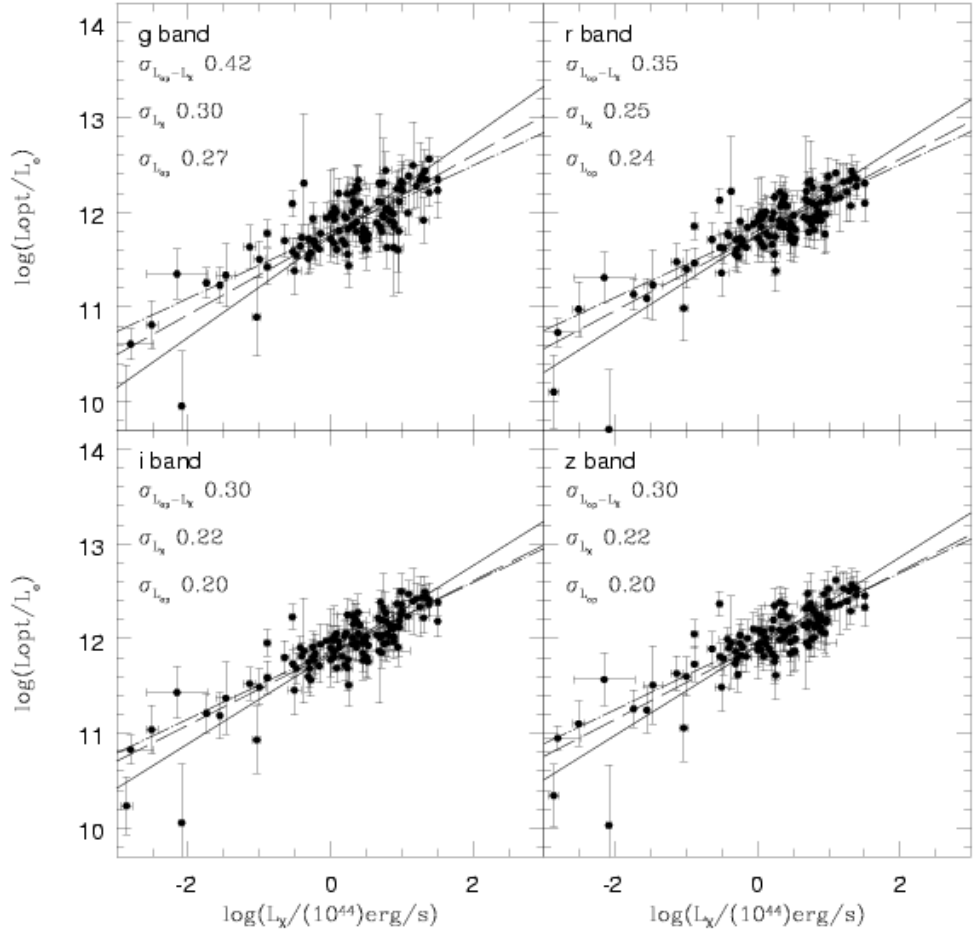}}
\end{minipage}
\end{center}
\caption{
Correlation between optical luminosities  and X-ray luminosities.  The
fit    is   performed   with     a     linear   regression in      the
$log(L_{opt})-log(L_X)$  space for each   of the 4 optical bands.  The
solid   and the dashed  lines  are the results  of  the orthogonal and
bisector  method  respectively over the  all  RASS-SDSS galaxy cluster
sample. The dot-dashed  line is the best  fit result of the orthogonal
method applied to the subsample of strictly X-ray selected REFLEX-NORAS
clusters. The error bars are at the 68\%  confidence level in both the
variables.}
\label{lx}
\end{figure}

\begin{figure}
\begin{center}
\begin{minipage}{1.0\textwidth}
\resizebox{\hsize}{!}{\includegraphics{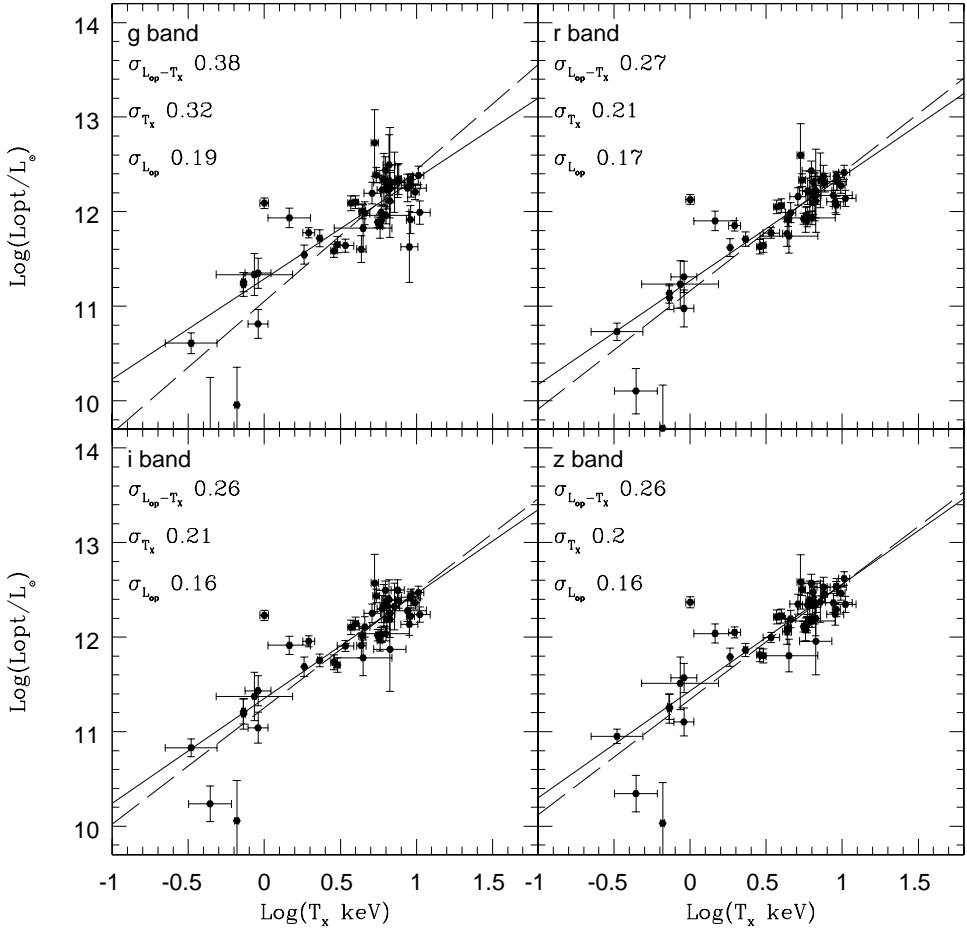}}
\end{minipage}
\end{center}
\caption{
Correlation between optical luminosities and ICM temperature.  The fit
is  performed with a  linear regression in the $log(L_{opt}-log(T_X))$
space for each of the 4 optical bands. The solid  and the dashed lines
are     the   results   of  the     orthogonal    and bisector  method
respectively. The error bars are at the 68\%  confidence level in both
the variables.}  
\label{lt}
\end{figure}

Figg. \ref{lx} and \ref{lt} show the $L_{op}  - L_X$ and $L_{op} - T_x
$ relation respectively, at the   radius of minimum scatter,  0.5   Mpc
$\rm{h}_{70}^{-1}$, in the z band. In fact the i  and the z bands have
a slightly  smaller scatter  than in the  other  optical bands at  any
radius.  Both plots show, as  an outliers, the cluster RXCJ0845.3+4430
, which features a deviation larger than $3 \sigma$  from the best fit.
The system is   a nearby group with   a  density contrast  too  low to
estimate  the optical luminosity reliably.    The X-ray luminosity  of
this system  has a 100\% error  respectively.   In the $L_{op}  - T_x$
there is another outlier:  the cluster RXCJ1629.6+4049. This system is
not a source of scatter  in the $L_{op}  - L_X$ relation and the error
in  the    X-ray   and   optical   luminosities  is    8\%   and  35\%
respectively. This can suggests  that the  optical luminosity is  well
measured, and it questions the estimate of  the temperature.  In fact,
the  X-ray  luminosity  of RXCJ1629.6+4049   is  2.78$\times  10^{43}$
ergs$\rm{s}^{-1}$,  and   the temperature,  estimated  from Horner  et
al. (2001), is 1 keV, while the $L_X-T_X$ relation of Xue
\& Wu, (2000), predicts at least a $T_X$ of 4.3 keV at that $L_X$.

With the $\sigma$  clipping  method those clusters were  excluded from
the  estimation  of the best   fit.  The best  fit   parameters of the
orthogonal and bisector methods, in the i and z photometric bands, are
shown  in the Table  \ref{lt1} and  \ref{bol}  for the $L_{op} -  L_X$
relation  for the   ROSAT X-ray luminosity   and  the bolometric X-ray
luminosity,  respectively. Table \ref{lt2}  shows the same results for
the $L_{op} - T_x $ relation. Table \ref{lt3} provides the results for
the $L_{op}-L_X$ relation in  i and z band  for the subsample of X-ray
selected REFLEX-NORAS clusters.    The tables show  also the estimated
orthogonal  scatter and the estimated  scatter  in both the variables.
The best fit  in the z band  for the $L_{op}  - L_X$ and the $L_{op} -
T_x $ relations at the radius of minimum scatter for all the RASS-SDSS
galaxy cluster sample are respectively:
\begin{equation}
L_{op}/L_{\odot}=10^{11.79 \pm 0.02} L_X(ROSAT)^{0.45\pm 0.03}
\label{eq1}
\end{equation}
\begin{equation}
L_{op}/L_{\odot}=10^{11.75 \pm 0.02} L_X(Bol)^{0.38\pm 0.02}
\label{eq2}
\end{equation}
\begin{equation}
L_{op}/L_{\odot}=10^{11.42 \pm 0.06} T_X^{1.12\pm 0.08}
\label{eq3}
\end{equation}
The value of   the  exponent in the   power  law for the  $L_{opt}   -
L_X(Bol)$ relation,   is  around 0.38   in the  region  of minimum, as
indicated in table \ref{lt1}. The values are not consistent within the
errors  with  the  value  of 0.5 predicted    under  the assumption of
hydrostatic equilibrium  and  constant mass  to light ratio.  The same
conclusion can  be derived for the  $L_{opt} - T_X$ relation   and from
the  $L_{op} - L_X$  for the subsample  of X-ray selected REFLEX-NORAS
clusters. A   simple reason of  the disagreement  could  be due to the
assumption of a constant mass to light ratio.  In fact, Girardi et al.
(2002) analysed in detail the mass to  light ratio in the  B band of a
sample of  294 clusters and groups ,  finding $M/L  \propto L^{0.33\pm
0.03}$. The same  results was  found by Lin  et al.  (2003) in  the  K
band. Thus if  we consider this  dependence of $M/L$  from the optical
luminosity  with the assumptions  of  hydrostatic equilibrium, the new
expected    relation between  the  optical   luminosity  and the X-ray
luminosity and temperature are $L_{op} \propto L_X^{0.4}$ and $L_{op}
\propto T_X^{1.25}$, respectively, which are in good agreement with
our results.

The scatter in the $L_{op} - L_x $ relation for  the aperture with the
best   correlation  ( 0.5  Mpc  $\rm{h}_{70}^{-1}$),  in  the  $L_{op}$
variable is 0.20, and the scatter in the $L_X$ variable is 0.22 in the
correlations obtained in   i and the  z  bands as  shown in  the Table
\ref{lt1}.  Therefore, by  calculating the total cluster luminosity in
the central part  of the system,  one can use  the $i$ or $z$  band to
predict the X-ray luminosity  from the optical  data with a mean error
of $60\%$.  In the same way the optical luminosity can be derived from
$L_X$  with the same uncertainty.  As indicated in the table \ref{lt3}
the uncertainty in the prediction  of  the two variables decreases  to
less  than  40\% if the   correlation is limited   to the REFLEX-NORAS
cluster subsample.  Table
\ref{lt2} shows that analogous results are obtained for the $L_{op}
- T_x $ relation. 

 Since the observational uncertainties in the  optical and in the X-ray
luminosity are about  20\%, the scatters  of 60\% of the overall sample in
both relations and of  40\% in the  the REFLEX-NORAS cluster subsample
for the $L_{op} - L_x $ relation should be intrinsic.

\begin{table}
\begin{tabular}[b]{c|ccccc|ccccc}
\multicolumn{11}{c}{$L_{op}-L_X(ROSAT)$ relation in the I band} \\ \hline
\multicolumn{6}{c|}{Orthogonal method}& \multicolumn{5}{|c}{Bisector method} \\ \hline
\renewcommand{\arraystretch}{0.2}\renewcommand{\tabcolsep}{0.05cm}
$r$ & $\alpha $ & $\beta$ & $\sigma$& $\sigma _{L_X}$ & $\sigma _{L_{op}}$ & $\alpha $ & $\beta$ & $\sigma$ &$\sigma _{L_X}$ & $\sigma _{L_{op}}$ \\
\hline
0.2 & $0.50 \pm 0.04$ & $11.41 \pm 0.03$ & 0.43& 0.32 & 0.30 & $0.32 \pm 0.01 $ & 1$1.44 \pm 0.03 $& 0.36 &0.23  &  0.28 \\
0.3 & $0.46 \pm 0.03$ & $11.70 \pm 0.02$ & 0.31& 0.23 & 0.21 & $0.36 \pm 0.01 $ & 1$1.71 \pm 0.02 $& 0.29 &0.18  &  0.23 \\
0.5 & $0.47 \pm 0.03$ & $11.83 \pm 0.02$ & 0.30& 0.22 & 0.20 & $0.38 \pm 0.01 $ & 1$1.85 \pm 0.02 $& 0.27 &0.18  &  0.20 \\
0.7 & $0.56 \pm 0.03$ & $11.94 \pm 0.03$ & 0.36& 0.26 & 0.25 & $0.47 \pm 0.01 $ & 1$1.96 \pm 0.02 $& 0.32 &0.22  &  0.24 \\
0.8 & $0.61 \pm 0.04$ & $11.99 \pm 0.03$ & 0.40& 0.29 & 0.28 & $0.51 \pm 0.01 $ & 1$2.01 \pm 0.03 $& 0.36 &0.24  &  0.26 \\
1.0 & $0.65 \pm 0.04$ & $12.05 \pm 0.03$ & 0.44& 0.31 & 0.31 & $0.54 \pm 0.01 $ & 1$2.07 \pm 0.03 $& 0.39 &0.27  &  0.28 \\
\hline								   							   
\end{tabular}	

\begin{tabular}[b]{c|ccccc|ccccc}
\multicolumn{11}{c}{$L_{op}-L_X(ROSAT)$ relation in the Z band} \\ \hline
\multicolumn{6}{c|}{Orthogonal method}& \multicolumn{5}{|c}{Bisector method} \\ \hline
\renewcommand{\arraystretch}{0.2}\renewcommand{\tabcolsep}{0.05cm}
$r$ & $\alpha $ & $\beta$ & $\sigma$& $\sigma _{L_X}$ & $\sigma _{L_{op}}$ & $\alpha $ & $\beta$ &$\sigma$& $\sigma _{L_X}$ & $\sigma _{L_{op}}$\\
\hline
0.2  & $0.50 \pm 0.04$& $11.50 \pm 0.03$& 0.43& 0.31 & 0.31& $0.31\pm 0.01$ & $11.54 \pm 0.03$ &  0.35 & 0.23& 0.26\\
0.3  & $0.45 \pm 0.03$& $11.79 \pm 0.02$& 0.31& 0.23 & 0.21& $0.37\pm 0.01$ & $11.80 \pm 0.02$ &  0.29 & 0.18& 0.23\\
0.5  & $0.47 \pm 0.03$& $11.92 \pm 0.02$& 0.30& 0.22 & 0.20& $0.39\pm 0.01$ & $11.93 \pm 0.02$ &  0.28 & 0.18& 0.21\\
0.7  & $0.56 \pm 0.03$& $12.03 \pm 0.03$& 0.36& 0.26 & 0.25& $0.47\pm 0.01$ & $12.04 \pm 0.02$ &  0.33 & 0.22& 0.24\\ 
0.8  & $0.60 \pm 0.04$& $12.08 \pm 0.03$& 0.40& 0.29 & 0.27& $0.51\pm 0.01$ & $12.10 \pm 0.03$ &  0.36 & 0.24& 0.26\\ 
1.0  & $0.65 \pm 0.04$& $12.14 \pm 0.03$& 0.44& 0.31 & 0.31& $0.54\pm 0.01$ & $12.15 \pm 0.03$ &  0.39 & 0.27& 0.28 \\
\hline
\end{tabular}								   
\caption{
The table presents the results of the best fit for the i and z band in
the region   of   the minimum  scatter, $0.2    \le  r  \le  1.0$  Mpc
$\rm{h}_{70}^{-1}$ for  the $L_{op}-L_X(ROSAT)$  relation. We show the
results    for the   two  methods  applied:   the orthogonal  distance
regression  (ODRPACK)   and the bisector method,  which    is the line
bisecting    the  best fit  results   of  the vertical  and horizontal
minimization.  The   $\alpha$ and  $\beta$   parameters are  given for
several apertures with  the 95\%   confidence errors.  The   orthogonal
scatter and the  scatters   in    log($L_X$(0.1-2.4 keV)    and  in
log($L_{op}$)   to the best fit  line   are given by $\sigma$, $\sigma
_{L_X}$ and $\sigma _{L_{op}}$ respectively.}
\label{lt1}							   
\end{table}	

\begin{table}
\begin{tabular}[b]{c|ccccc|ccccc}
\multicolumn{11}{c}{$L_{op}-L_X(Bolometric)$ relation in the I band} \\ \hline
\multicolumn{6}{c|}{Orthogonal method}& \multicolumn{5}{|c}{Bisector method} \\ \hline
\renewcommand{\arraystretch}{0.2}\renewcommand{\tabcolsep}{0.05cm}
$r$ & $\alpha $ & $\beta$ & $\sigma$ &$\sigma _{L_X}$ & $\sigma
_{L_{op}}$ & $\alpha $ & $\beta$ &$\sigma$& $\sigma _{L_X}$ & $\sigma
_{L_{op}}$\\
\hline
0.2&  $0.38\pm 0.03$ &  $11.25 \pm  0.04$ & 0.41 &  0.31 & 0.27 & $0.27 \pm 0.01$ &  $11.31\pm  0.03$ & 0.37 & 0.24 & 0.28 \\
0.3&  $0.36\pm 0.02$ &  $11.54 \pm  0.03$ & 0.31 &  0.23 & 0.20 & $0.30 \pm 0.01$ &  $11.57\pm  0.02$ & 0.29 & 0.18 & 0.23 \\
0.5&  $0.38\pm 0.02$ &  $11.66 \pm  0.03$ & 0.30 &  0.22 & 0.20 & $0.32 \pm 0.01$ &  $11.70\pm  0.02$ & 0.27 & 0.18 & 0.21 \\
0.7&  $0.38\pm 0.02$ &  $11.66 \pm  0.03$ & 0.30 &  0.22 & 0.20 & $0.32 \pm 0.01$ &  $11.70\pm  0.02$ & 0.27 & 0.18 & 0.21 \\
0.8&  $0.49\pm 0.03$ &  $11.77 \pm  0.04$ & 0.41 &  0.30 & 0.28 & $0.42 \pm 0.01$ &  $11.82\pm  0.03$ & 0.37 & 0.25 & 0.27 \\
1.0&  $0.53\pm 0.03$ &  $11.81 \pm  0.04$ & 0.44 &  0.32 & 0.31 & $0.45 \pm 0.01$ &  $11.86\pm  0.03$ & 0.40 & 0.27 & 0.29 \\

\hline								   							   
\end{tabular}	

\begin{tabular}[b]{c|ccccc|ccccc}
\multicolumn{11}{c}{$L_{op}-L_X(Bolometric)$ relation in the Z band} \\ \hline
\multicolumn{6}{c|}{Orthogonal method}& \multicolumn{5}{|c}{Bisector method} \\ \hline
\renewcommand{\arraystretch}{0.2}\renewcommand{\tabcolsep}{0.05cm}
$r$ & $\alpha $ & $\beta$ & $\sigma$& $\sigma _{L_X}$ & $\sigma _{L_{op}}$ & $\alpha $ & $\beta$ &$\sigma$& $\sigma _{L_X}$ & $\sigma _{L_{op}}$\\
\hline

0.2&   $0.38\pm 0.03$ & $11.34 \pm 0.04$ & 0.41 & 0.30& 0.28 & $0.26\pm  0.01$ &$ 11.42 \pm  0.03$  & 0.35 &0.23 & 0.27 \\
0.3&   $0.35\pm 0.02$ & $11.63 \pm 0.03$ & 0.30 & 0.24& 0.19 & $0.31\pm  0.01$ &$ 11.65 \pm  0.02$  & 0.29 &0.18 & 0.23 \\
0.5&   $0.38\pm 0.02$ & $11.75 \pm 0.03$ & 0.30 & 0.23& 0.20 & $0.32\pm  0.01$ &$ 11.78 \pm  0.02$  & 0.28 &0.18 & 0.21 \\
0.7&   $0.38\pm 0.02$ & $11.75 \pm 0.03$ & 0.30 & 0.23& 0.20 & $0.32\pm  0.01$ &$ 11.78 \pm  0.02$  & 0.28 &0.18 & 0.21 \\
0.8&   $0.49\pm 0.03$ & $11.86 \pm 0.04$ & 0.40 & 0.29& 0.28 & $0.42\pm  0.01$ &$ 11.90 \pm  0.03$  & 0.37 &0.25 & 0.27 \\
1.0&   $0.52\pm 0.03$ & $11.90 \pm 0.04$ & 0.44 & 0.32& 0.31 & $0.44\pm  0.01$ &$ 11.95 \pm  0.03$  & 0.40 &0.28 & 0.29 \\

\hline
\end{tabular}								   
\caption{
The table presents the results of the best fit for the i and z band in
the   region   of the  minimum scatter,   $0.2   \le r  \le   1.0$ Mpc
$\rm{h}_{70}^{-1}$  for the $L_{op}-L_X(Bolometric)$ relation. We show
the results for   the two  methods  applied: the  orthogonal  distance
regression    (ODRPACK) and the  bisector method,    which is the line
bisecting  the best  fit   results  of  the vertical and    horizontal
minimization. The  $\alpha$  and $\beta$   parameters are  given   for
several  apertures  with  the  95\%  confidence errors. The  orthogonal
scatter and the scatters in  log($T_X$) and in log($L_{op}$) to the
best fit line  are  given by  $\sigma$,  $\sigma  _{T_X}$  and $\sigma
_{L_{op}}$ respectively.}
\label{bol}							   
\end{table}

\begin{table}
\begin{tabular}[b]{c|ccccc|ccccc}
\multicolumn{11}{c}{$L_{op}-T_X$ relation in the I band} \\ \hline
\multicolumn{6}{c|}{Orthogonal method}& \multicolumn{5}{|c}{Bisector method} \\ \hline
\renewcommand{\arraystretch}{0.2}\renewcommand{\tabcolsep}{0.05cm}
$r$ & $\alpha $ & $\beta$ & $\sigma$ &$\sigma _{T_X}$ & $\sigma
_{L_{op}}$ & $\alpha $ & $\beta$ &$\sigma$& $\sigma _{T_X}$ & $\sigma
_{L_{op}}$\\
\hline
 0.2 & $ 0.97\pm 0.17$ & $11.05 \pm 0.12$ & 0.53 & 0.39 & 0.36 &$1.22 \pm 0.22$ & $10.80 \pm  0.16$ & 0.54 & 0.39 & 0.37 \\
 0.3 & $ 1.06\pm 0.09$ & $11.21 \pm 0.06$ & 0.32 & 0.27 & 0.17 &$1.17 \pm 0.18$ & $11.12 \pm  0.14$ & 0.31 & 0.18 & 0.25 \\
 0.5 & $ 1.11\pm 0.08$ & $11.33 \pm 0.06$ & 0.26 & 0.20 & 0.16 &$1.13 \pm 0.10$ & $11.30 \pm  0.08$ & 0.25 & 0.17 & 0.19 \\
 0.7 & $ 1.20\pm 0.11$ & $11.43 \pm 0.08$ & 0.30 & 0.23 & 0.19 &$1.40 \pm 0.12$ & $11.27 \pm  0.09$ & 0.29 & 0.20 & 0.21 \\
 0.8 & $ 1.20\pm 0.12$ & $11.51 \pm 0.09$ & 0.33 & 0.26 & 0.20 &$1.51 \pm 0.13$ & $11.26 \pm  0.10$ & 0.33 & 0.23 & 0.24 \\
 1.0 & $ 1.21\pm 0.14$ & $11.60 \pm 0.10$ & 0.39 & 0.32 & 0.23 &$1.62 \pm 0.16$ & $11.26 \pm  0.12$ & 0.39 & 0.27 & 0.28 \\

\hline								   							   
\end{tabular}	

\begin{tabular}[b]{c|ccccc|ccccc}
\multicolumn{11}{c}{$L_{op}-T_X$ relation in the Z band} \\ \hline
\multicolumn{6}{c|}{Orthogonal method}& \multicolumn{5}{|c}{Bisector method} \\ \hline
\renewcommand{\arraystretch}{0.2}\renewcommand{\tabcolsep}{0.05cm}
$r$ & $\alpha $ & $\beta$ & $\sigma$& $\sigma _{T_X}$ & $\sigma _{L_{op}}$ & $\alpha $ & $\beta$ &$\sigma$& $\sigma _{T_X}$ & $\sigma _{L_{op}}$\\
\hline

0.2 &  $0.97 \pm 0.18$ & $11.18 \pm 0.13$ & 0.54 & 0.37 & 0.40 & $1.16 \pm 0.18$ & $10.95 \pm 0.13$ & 0.55 & 0.42 & 0.35 \\
0.3 &  $1.06 \pm 0.08$ & $11.31 \pm 0.06$ & 0.31 & 0.27 & 0.16 & $1.18 \pm 0.19$ & $11.20 \pm 0.15$ & 0.31 & 0.18 & 0.25 \\
0.5 &  $1.12 \pm 0.08$ & $11.42 \pm 0.06$ & 0.25 & 0.20 & 0.16 & $1.12 \pm 0.10$ & $11.39 \pm 0.08$ & 0.25 & 0.16 & 0.19 \\
0.7 &  $1.20 \pm 0.11$ & $11.52 \pm 0.08$ & 0.30 & 0.24 & 0.19 & $1.40 \pm 0.12$ & $11.34 \pm 0.09$ & 0.30 & 0.20 & 0.23 \\
0.8 &  $1.21 \pm 0.12$ & $11.60 \pm 0.09$ & 0.34 & 0.27 & 0.20 & $1.51 \pm 0.13$ & $11.34 \pm 0.10$ & 0.33 & 0.23 & 0.24 \\
1.0 &  $1.22 \pm 0.14$ & $11.68 \pm 0.10$ & 0.39 & 0.32 & 0.23 & $1.62 \pm 0.15$ & $11.34 \pm 0.11$ & 0.39 & 0.27 & 0.28 \\

\hline
\end{tabular}								   
\caption{
The table presents the results of the best fit for the i and z band in
the region   of  the  minimum  scatter, $0.2    \le r  \le   1.0$  Mpc
$\rm{h}_{70}^{-1}$.  We show the results  for the two methods applied:
the orthogonal distance  regression (ODRPACK) and the bisector method,
which is the line bisecting the  best fit results  of the vertical and
horizontal minimization. The $\alpha$ and $\beta$ parameters are given
for several apertures with  the 95\% confidence errors. The orthogonal
scatter and the scatters in  log($T_X$) and in log($L_{op}$) to the
best fit  line  are  given by $\sigma$,  $\sigma  _{T_X}$  and $\sigma
_{L_{op}}$ respectively.}
\label{lt2}							   
\end{table}

\begin{table}
\begin{tabular}[b]{c|ccccc|ccccc}
\multicolumn{11}{c}{$L_{op}-L_X(ROSAT)$ relation for REFLEX-NORAS clusters only} \\ \hline
\multicolumn{11}{c}{orthogonal method} \\ \hline
\multicolumn{6}{c|}{i band}& \multicolumn{5}{|c}{z band} \\ \hline
\renewcommand{\arraystretch}{0.2}\renewcommand{\tabcolsep}{0.05cm}
$r$ & $\alpha $ & $\beta$ & $\sigma$& $\sigma _{L_X}$ & $\sigma _{L_{op}}$ & $\alpha $ & $\beta$ & $\sigma$ &$\sigma _{L_X}$ & $\sigma _{L_{op}}$ \\
\hline
0.2 & $0.28 \pm 0.04$ & $11.48 \pm 0.03$ & 0.32& 0.23 & 0.22 & $0.27 \pm 0.04 $ & $11.57 \pm 0.03 $& 0.31 &0.23  &  0.21 \\
0.3 & $0.33 \pm 0.03$ & $11.74 \pm 0.02$ & 0.23& 0.16 & 0.15 & $0.34 \pm 0.03 $ & $11.83 \pm 0.02 $& 0.21 &0.16  &  0.15 \\
0.5 & $0.36 \pm 0.03$ & $11.87 \pm 0.02$ & 0.22& 0.16 & 0.15 & $0.36 \pm 0.03 $ & $11.97 \pm 0.02 $& 0.22 &0.16  &  0.14 \\
0.7 & $0.36 \pm 0.03$ & $12.01 \pm 0.02$ & 0.24& 0.18 & 0.16 & $0.37 \pm 0.03 $ & $12.11 \pm 0.02 $& 0.24 &0.17  &  0.16 \\
0.8 & $0.37 \pm 0.03$ & $12.08 \pm 0.03$ & 0.27& 0.20 & 0.17 & $0.37 \pm 0.03 $ & $12.17 \pm 0.02 $& 0.26 &0.20  &  0.17 \\
1.0 & $0.37 \pm 0.04$ & $12.18 \pm 0.03$ & 0.29& 0.23 & 0.18 & $0.37 \pm 0.04 $ & $12.28 \pm 0.03 $& 0.28 &0.22  &  0.18 \\
\hline								   							   
\end{tabular}	
							   
\caption{
The table presents the results of the best fit for the i and z band in
the region of    the  minimum scatter,   $0.2   \le r \le    1.0$  Mpc
$\rm{h}_{70}^{-1}$   for  the    $L_{op}-L_X(ROSAT)$ relation  in  the
subsample of  REFLEX and NORAS   X-ray selected clusters.  We  show the
results for the orthogonal distance regression  (ODRPACK) method.  The
$\alpha$ and $\beta$ parameters are  given for several apertures  with
the 95\%  confidence errors.  The  orthogonal scatter and the scatters
in  log($L_X$(0.1-2.4 keV))  and in  log($L_{op}$)  to the  best fit
line  are  given by $\sigma$,  $\sigma  _{L_X}$ and $\sigma _{L_{op}}$
respectively.}
\label{lt3}							   
\end{table}

\section{Summary and Conclusions}
We created a  database of clusters  of galaxies  based on  the largest
available X-ray  and optical surveys: the ROSAT  All Sky Survey (RASS)
and the Sloan Digital Sky Survey (SDSS).  The RASS-SDSS galaxy cluster
catalog is the first catalog which combines X-ray and optical data for
a  large   number (114) of  galaxy clusters.    The systems  are X-ray
selected, ranging from  groups of
 $10^{12.5}$  $M_{\odot}$ to massive
clusters of  $10^{15}$ $M_{\odot}$
 in  a redshift range from 0.002 to
0.45.    The X-ray  (luminosity     in the   ROSAT band,
   bolometric
luminosity,  redshift, center  coordinates)  and   optical  properties
(Schechter luminosity function
 parameters, luminosity, central galaxy
density, core, total and
 half-light radii) are  computed in a uniform
and accurate  way.  The catalog   contains also temperature  and  iron
abundance for a subsample of 53 clusters from the Asca Cluster Catalog
and the Group Sample. The resulting RASS-SDSS galaxy cluster catalog,
then,  constitutes an  important database to  study  the  properties of
galaxy clusters   and  in   particular the  relation   of  the  galaxy
population seen in the optical to the properties of the X-ray luminous
ICM.

The  first  investigations reported  have  shown  a tight  correlation
between the X-ray  and optical   properties, when  the choice  of  the
measurement   aperture  for the optical     luminosity  and the  optical
wavelength band  are optimized.  We  found that the optical luminosity
calculated in the i and in the z band correlates better with the X-ray
luminosity and the  ICM temperature, in  comparison to the other Sloan
photometric   bands.  Thus the    red optical  bands,  which are  more
sensitive to the light of the  old stellar population and therefore to
the stellar mass of cluster galaxies, have tight correlations with the
X-ray properties of the systems. 

Moreover,  we   found  that the   scatter  in   the  $L_{op}-L_X$  and
$L_{op}-T_X$ relations can be minimized  if the optical luminosity  is
measured   within   a    cluster    aperture  between    0.2-0.8   Mpc
$\rm{h}_{70}^{-1}$, with an absolute minimum of the scatter at 0.5 Mpc
$\rm{h}_{70}^{-1}$. The best  aperture   in the  central  part of  the
cluster for the  measurement of the optical luminosity  is due  to the
fact that it is  a good compromise to  simultaneously assess the total
richness and the compactness of the cluster.

Finally   by  using the   relations obtained   from   the  z band,  we
demonstrated that, given  the optical properties  of a cluster, we can
predict the X-ray luminosity and temperature with  an accuracy of 60\%
and  vice versa.  By    restricting  the correlation  analysis  to  the
subsample of X-ray detected REFLEX-NORAS clusters, the minimum scatter
decreases to less than 40\%  for the $L_{op}-L_X$ relation.  Since the
observational uncertainties in the optical and in the X-ray luminosity
are   about 20\%, the observed scatters   in both  relations should be
intrinsic.

The resulting logarithmic slope for the $L_{op}-L_X$ relation with the
minimum  scatter    is   $0.38\pm 0.02$,  while   the   value  for the
$L_{op}-T_X$  relation  is $1.12  \pm  0.08$.   Both results  are  not
consistent  with   the assumption of  hydrostatic  equilibrium   and a
constant  M/L.  If we  assume that M/L depends  on the luminosity with
the power law $M/L \propto L^{0.3}$ (Girardi et al. 2002), our results
are  in very   good   agreement with  the  expected values   under the
assumption of hydrostatic equilibrium.

 The analysis carried out in this paper on the correlation between
X-ray     and optical  apparence  of  galaxy    clusters is completely
empirical. In principle, the best way to proceed in this kind of study
is  to measure the optical luminosity  within the physical size of the
cluster, like the virial radius. Without optical spectroscopic data or
accurate  temperature measurements, the  cluster virial  radius can be
calculated by assuming  a    theoretical model relating   the  optical
luminosity to the  cluster   mass.  At this stage   of  the  work,  we
preferred  to tackle the  cluster   X-ray-optical connection with  the
empirical method   explained    in the    paper,  in  order   to  have
model-independent results. On the other hand,  not taking into account
the different cluster sizes could have affected both the slope and the
scatter of the given relations (eq.  \ref{eq1}, \ref{eq2} and
\ref{eq3}).  Therefore,  for a better   understanding of the important
connection between the X-ray and optical apparence of galaxy clusters,
the optical luminosity has to  be calculated within the physical  size
of the cluster.  This work is in progress and will be published in the
second  paper   of this  series about   the  RASS-SDSS  galaxy cluster
sample. The next step will be the  study of the fundamental plane of
galaxy clusters. Through this kind of analysis we will find out if the
observed scatter  in the correlations   between the optical  and X-ray
properties  depends  on  another  parameter   related to the   cluster
compactness.  Moreover, because of the link between the galaxy cluster
fundamental plane and the M/L parameter,  we will connect directly the
slope of two relations to the behavior of M/L.

\vspace{2cm}

Funding for the creation and distribution of the SDSS Archive has been
provided  by    the  Alfred P.   Sloan   Foundation, the Participating
Institutions, the  National Aeronautics and  Space Administration, the
National  Science  Foundation, the  U.S.    Department of  Energy, the
Japanese Monbukagakusho, and the Max Planck Society. The SDSS Web site
is   http://www.sdss.org/. The  SDSS is  managed  by the Astrophysical
Research Consortium   (ARC) for the   Participating Institutions.  The
Participating Institutions  are The  University of Chicago,  Fermilab,
the Institute for Advanced  Study, the Japan Participation  Group, The
Johns  Hopkins    University,  Los Alamos   National  Laboratory,  the
Max-Planck-Institute for   Astronomy (MPIA), the  Max-Planck-Institute
for   Astrophysics (MPA), New Mexico  State  University, University of
Pittsburgh, Princeton University, the United States Naval Observatory,
and the University of Washington.

\section{References}
 Abazajian, K., Adelman, J., Agueros, M.,et al. 2003, AJ, 126, 2081 (Data Release One)\\
 Adami, C. et al. 1998, A\&A, 336, 63;
\\
 Akritas, M., Mazure, A., Katgert, P., et al. 1996, ApJ, 470, 706;
\\
 Arnaud, M., Rothenflug, R.; Boulade, O., et al. 1992, A\&A, 254, 49;
\\
 Bahcall, N. 1977, ApJL, 217, 77;
\\
 Bahcall, N. 1981, ApJ, 247, 787;
\\
 Bahcall, N., McKay, T. A., Annis, J., et al. 2003, ApJS, 148, 253;
\\
 Blanton, M., Blanton, M. R., Dalcanton, J., Eisenstein, D., et al. 2001, AJ, 121, 2358;
\\
 Blanton, M.R., Lupton, R.H., Maley, F.M. et al. 2003, AJ, 125, 2276 (Tiling Algorithm)\\
 B\"ohringer, H., Voges, W.; Huchra, J. P., et al. 2000, ApJS, 129, 435;
\\
 B\"ohringer, H.,  Schuecker, P., Guzzo, L., et al. 2001, A\&A, 369, 826;
\\
 B\"ohringer, H., Collins, C. A., Guzzo, L., et al. 2002, ApJ, 566, 93;
\\
 De grandi, S. and  Molendi, S. 2002, ApJ, 567, 163;
\\
 De Propris, R., Colless, M., Driver, S. P., et al. 2003, MNRAS, 342, 725;
\\
 Donahue, M., Mack, J., Scharf, C., et al. 2001, ApJL, 552, 93;
\\
 Donahue, M., Scharf, C. A., Mack, J., et al. 2002, ApJ, 569, 689;
\\
 Dressler, A., Oemler, A. Jr., Couch, W. J., et al. 1997, ApJ, 490, 577;
\\
 Edge, A. C. \& Stewart, G. C.  1991, MNRAS, 252, 428;
\\
 Eisenstein, D. J., Annis, J., Gunn, J. E., et al. 2001, AJ, 122, 2267;
\\
 Fasano, G., Poggianti, B. M., Couch, W. J., et al. 2000, ApJ, 542, 673;
\\
 Finoguenov, A., Arnaud, M., David, L. P. 2001, ApJ, 555, 191F;
\\
 Finoguenov, A., Reiprich, T. H., Böhringer, H. 2001,A\&A,368,749;\\
 Flectcher, R. 1970, Comput. J, 13, 317;
\\
 Fukugita, M., Shimasaku, K.; Ichikawa, T. 1995, PASJ, 107,945;
\\
 Fukugita, M., Ichikawa, T., Gunn, J. E. 1996, AJ, 111, 1748;
\\
 Girardi, M., Borgani, S., Giuricin, G., et al. 2000, ApJ, 530, 62;
\\
 Girardi, M., Manzato, P., Mezzetti, M., et al. 2002, ApJ, 569, 720;
\\
 Gladders, M. and Yee, H. K. C. 2000, AJ, 120, 2148;
\\
 Gomez, P., Nichol, R. C., Miller, C. J., et al. 2003, ApJ, 584, 210;
\\
 Goto, T., Sekiguchi, M., Nichol, R. C., et al. 2002, AJ, 123, 1807;
\\
 Goto, T., Okamura, S., McKay, T. A.,  et al. 2002, PASP, 123, 1807;
\\
 Gunn, J.E., Carr, M.A., Rockosi, C.M., et al 1998, AJ, 116, 3040 (SDSS Camera)\\
 Hogg, D.W., Finkbeiner, D. P., Schlegel, D. J., Gunn, J. E. 2001, AJ, 122, 2129;\\
 Horner, D. 2001, PhD Thesis, University of Maryland;
\\
 Ikebe, Y., Reiprich, T. H., B\"ohringer, H., et al. 2002, A\&A, 383, 773;
\\
 Isobe, T., Feigelson, E. D., Akritas, M. G., et al. 1990, ApJ, 364, 104;
\\
 Kelson, D., van Dokkum, P. G., Franx, M., et al. 1997, ApJ, 478, 13;
\\
 Kelson, D., llingworth, G. D., van Dokkum, P. G., Franx, M. 2000, ApJ, 531, 137;
\\
 Kim, R., Seung J., Kepner, J. V. 2002, AJ, 123, 20;
 Kochanek, C. S., Pahre, M. A., Falco, E. E., et al. 2001, ApJ, 560,566;
\\
 Lin, Y., Mohr, J. J., Stanford, S. A. 2003, ApJ,591,749;
\\
 Lubin, L., Oke, J. B.; Postman, M. 2002, AJ, 124, 1905;
\\
 Lumsden, S. L.,  Collins, C. A., Nichol, R. C.,et al. 1997, MNRAS, 290, 119;
\\
 Lupton, R. H., Gunn, J. E., Szalay, A. S. 1999, AJ, 118, 1406;
\\
 Lupton, R., Gunn, J. E., Ivezi\'c, Z.,  et al.  2001, in ASP Conf. Ser. 238, 
Astronomical Data Analysis Software and Systems
     X, ed. F. R. Harnden, Jr., F. A. Primini, 
and H. E. Payne (San Francisco: Astr. Soc. Pac.), p. 269 (astro-ph/0101420);\\
 Markevitch, M.  1998, ApJ, 504, 27;
\\
 Mulchaey, J.S., Davis, D. S., Mushotzky, R. F.; Burstein, D. 2003,ApJSS, 145, 39;
\\
 Pier, J.R., Munn, J.A., Hindsley, R.B., et al. 2003, AJ, 125, 1559 (Astrometry)\\
 Poggianti, B. et al.,  Bridges, T.J., Komiyama, Y., 2003, astro-ph/0309449;
\\
 Postman, M., Lubin, L. M., Oke, J. B. 1998, AJ, 116, 560;
\\
 Reiprich, T.H. and B\"oringher H. 2002, ApJ, 567, 716.
\\
 Retzlaff, J. 2001,XXIst Moriond Astrophysics Meeting, 
March 10-17, 2001 Savoie, France. Edited by D.M. Neumann  J.T.T. Van.\\
 Rosati, P., Borgani S., Norman, C. 2002, ARAA, 40, 539;
\\
 Sarazin, C. 1980, ApJ, 236, 75;
\\
 Schlegel, D., Finkbeiner, D. P., Davis, M. 1998, ApJ, 500, 525;
\\
 Shimasaku, K., Fukugita, M., Doi, M., et al. 2001, AJ, 122, 1238;
\\
 Smith, J.A., Tucker, D.L., Kent, S.M., et al. 2002, AJ, 123, 2121;\\
 Stoughton, C., Lupton, R.H., Bernardi, M., et al. 2002, AJ, 123, 485;
\\
 Strateva, I., Ivezi\'c, Z., Knapp, G. R., et al. 2001, AJ, 122, 1861;
\\
 Strauss, M. A., M.A., Weinberg, D.H., Lupton, R.H. et al. 2002, AJ, 124, 1810;
\\
 Van Dokkum, S., Franx, M., Fabricant, D., et al. 2000, ApJ, 541, 95V;
\\
 Voges, W., Aschenbach, B., Boller, Th., et al. 1999, A\&A, 349, 389;
\\
 Xue, Y. \& Wu, X. 2000, ApJ, 538, 65;
\\
 Yasuda, N., Fukugita, M. Narayanan, V. K. et al. 2001, AJ, 122, 1104;
\\
 Yee, YH. K. C. and Ellingson, E.  2003, ApJ, 585, 215;
\\
 York, D. G., Adelman, J., Anderson, J.E.,  et al. 2000, AJ, 120, 1579;
\\
 Ziegler, B. \&Bender, R. 1997, MNRAS, 291, 527;
\\

\clearpage

\clearpage

\clearpage

\clearpage

\clearpage

\clearpage

\clearpage

\clearpage

\clearpage

\clearpage

\clearpage

\clearpage

\clearpage

\clearpage

\clearpage

\clearpage

\clearpage

\clearpage

\clearpage

\end{document}